# Discovery of Electrochemically Induced Grain Boundary Transitions

Jiuyuan Nie [1,#], Chongze Hu [1,2,#], Qizhang Yan [1], and Jian Luo [1,2,*]

[1] Department of Nanoengineering; [2] Program of Materials Science and Engineering,

University of California, San Diego,

La Jolla, California 92093, U.S.A.

[#] These authors contributed equally.

[*] Corresponding author. E-mail address: jluo@alum.mit.edu (J. Luo).

**Abstract**

Electric fields and currents, which are used in innovative materials processing and electrochemical energy conversion, can often alter microstructures in unexpected ways. However, little is known about the underlying mechanisms. Using $ZnO$-$Bi_2O_3$ as a model system, this study uncovers how an applied electric current can change the microstructural evolution through an electrochemically induced grain boundary (GB) transition. By combining aberration-corrected electron microscopy, photoluminescence spectroscopy, first-principles calculations, a generalizable thermodynamic model, and *ab initio* molecular dynamics, this study reveals that electrochemical reduction can cause a GB disorder-to-order transition to markedly increase GB diffusivities and mobilities. Consequently, abruptly enhanced or abnormal grain growth takes place. These findings advance our fundamental knowledge of GB complexion (phase-like) transitions and electric field effects on microstructural stability and evolution, with broad scientific and technological impacts. A new method to tailor the GB structures and properties, as well as the microstructures, electrochemically can also be envisioned.

**Introduction**

It has been long proposed[1] that grain boundaries (GBs) can be considered as two-dimensional (2D) interfacial phases (*a.k.a.* "complexions"[2-7]), which can undergo transitions to influence various kinetic, mechanical, chemical, electronic, ionic, and other properties[3-5,7-12]. However, prior studies have seldom elucidated the mechanism of how an external stimulus – other than (a very few instances of) temperature or segregation – can induce a GB transition and subsequently alter the properties with clear underlying mechanisms. Notably, it was proposed that GB transitions can alter microstructural evolution abruptly.[3,4,7] Also, interestingly, electric fields and currents, which are used in various innovative materials processing[13-16] and electrochemical energy conversion[17,18] and storage[19] devices, can often alter microstructures unexpectedly and abruptly. Yet, the underlying atomic-level mechanisms remain elusive.

This study first aims at decoding how an electric field/current can alter microstructural evolution, an outstanding scientific problem of fundamental interest yet with broad technological implications. A spectrum of fascinating and intriguing observations of the "electric field effects" of suppressed[20,21] *vs.* enhanced[17,20-25] (including abnormal[17,23]) grain growth have been made in several oxides. See Supplementary Note 1 for elaboration, along with a discussion of relevant materials processing technologies[13-16] (including methods to sinter ceramics in seconds[13,16]) where electric fields/currents can affect microstructural evolution. Moreover, electric fields and currents are present in solid oxide fuel cells[17,18] solid-state batteries[19], and various other electrochemical and electronic devices, where they can cause unexpected (often undesirable or even catastrophic) changes in microstructures or GB properties.

In a broader context, the formation and transition of 2D interfacial phases,[1] which were also named as "complexions"[2-7] to differentiate them from thin layers of precipitated 3D bulk phases at GBs, can often control various materials properties.[3-5,7-12] However, the majority of prior studies focused on symmetric tilt or twist GBs that are relatively easy to image and model. For example, a most recent study observed GB phase transitions at a symmetric tilt GB in pure copper.[26] General GBs (asymmetric GBs that are often of mixed tilt and twist characters[27]) are much less understood, but they are ubiquitous and can often be the weaker links mechanically and chemically in polycrystalline materials.[9,11,12] Moreover, Dillon and Harmer proposed that anisotropic complexion transitions at general GBs can cause the AGG,[3,4,7] which shed light on one of the most long-standing mysteries in materials science but open questions remain. Specifically, how an abnormal grain can initiate remains under debate for nearly a century, albeit the cause can vary for different cases.

In this study, we use ZnO-$Bi_2O_3$ as a model system to uncover how an applied electric current can change the microstructural evolution (including triggering AGG) through electrochemically induced transitions at general GBs. Here, we combine aberration-corrected electron microscopy, first-principles calculations, and *ab initio* molecular dynamics to reveal that electrochemical reduction can cause GB disorder-to-order transitions to markedly increase GB diffusivities. Consequently, enhanced or abnormal grain growth takes place. A generalizable thermodynamic model is further proposed. This work builds a bridge between two important areas of GB complexion (phase-like) transitions and electric field effects on microstructural stability and evolution, while significantly advancing our fundamental knowledge in both areas. Furthermore, the discovery opens a new window to tailor a broad range of GB (*e.g.*, electronic or ionic) properties, as well as the microstructures, electrochemically.

**Results**

**Characteristic amorphous-like GBs in ZnO-$Bi_2O_3$.** Here, we select $Bi_2O_3$-doped ZnO (with 0.5 mol% $Bi_2O_3$ added in the polycrystal) as our model system. The solid solubility limit of $Bi_2O_3$ in the ZnO crystal is <0.06 mol% [28] so that most $Bi_2O_3$ is present at GBs and triple-grain junctions.[29] The formation of ~0.7-0.9 nm thick liquid-like intergranular films (IGFs) at general GBs is confirmed in our reference specimen without an applied electric field by aberration-corrected scanning and high-resolution transmission electron microscopy (AC STEM and HRTEM; Fig. 1a, b). Such nanometer-thick IGFs, which have been equivalently interpreted as equilibrium-thickness liquid-like interfacial films by Clarke[30] or disordered multilayer adsorbates by Cannon *et al.*,[31] represent one most widely observed complexion in ceramics.[4,5,8,29-31] These IGFs are also called "amorphous-like" or "disordered" GBs here, albeit the existence of partial orders.[5,6] Prior studies have demonstrated that such amorphous-like IGFs form at all general GBs in $Bi_2O_3$-saturated ZnO at thermodynamic equilibria both above and below the bulk eutectic temperature.[29] Surprisingly and interestingly, here we observe highly ordered GB structures in ZnO-$Bi_2O_3$ (Fig. 1c, d, g, h) that have never been reported before.

**Electrochemical reduction enhanced grain growth**. We design and fabricate a Polycrystal 1/Single Crystal/Polycrystal 2 (PC1/SC/PC2) sandwich specimen to conduct a grain growth experiment with a constant current density of 6.4 mA/mm$^2$ (Fig. 2a and Supplementary Fig. 1 for more details). First, we observe AGG in the PC1- region (where "-" *vs.* "+" denotes the reduced *vs.* oxidized region) near the negative electrode (cathode), as evident in Fig. 2b, c. Second, we observe enhanced migration of the SC/PC2- interface (Fig. 2f, g) *vs.* only moderate migration of the PC1+/SC interface (Fig. 2d, e). Quantitative measurements of the migration distances (shown in Supplementary Fig. 2) reveal abruptly enhanced migration at the (most reduced) middle section of the SC/PC2- interface, in comparison with the (oxidized) PC1+/SC interface and both interfaces in a reference sandwich specimen annealed with no electric field. A detailed and quantitative comparison can be found in Supplementary Note 2.

Next, let us show that the PC1- and PC2- regions in our sandwich specimen (Fig. 2), where we observe increased GB mobilities (either AGG or enhanced migration of the SC/PC interface), are oxygen reduced. Here, Bi$_2$O$_3$-enriched liquid-like IGFs in polycrystals are ion-conducting, while the ZnO single crystal is electron-conducting but ion-blocking (see Supplementary Note 3). Hence, PC1+ and PC2+ regions are oxidized (as the oxygen ions are blocked and accumulated near the interfaces), which is evident by the pore formation (see, *e.g.*, Fig. 2d) due to an oxidation reaction that produces O$_2$. On the other hand, the PC1- and PC2- regions must be reduced (due to the depletion of oxygen ions). Subsequently, we can sketch the profiles of electric ($\phi$), chemical ($\mu_{O^{2-}}$), and electrochemical ($\eta_{O^{2-}}$) potentials *vs.* locations in Fig. 2h according to a model presented in Supplementary Note 3. The electrochemical reduction in the PC1- region near the negative electrode is well expected. To prove the reduction in the PC2- region, we conduct spatially resolved photoluminescence spectroscopy to probe oxygen vacancies (Fig. 3a). The integrated photoluminescence intensities for the combined photoluminescence peak at ~400-700 nm (representing all defects at GBs) at different locations are shown in Fig. 3c, d. We further decompose this combined peak into a "green" band that is known to represent the oxygen vacancies and an "orange" band that most likely represents Zn interstitials (see Supplementary Note 4 and Supplementary Fig. 8). The decomposed green-band peaks at selected locations shown in Fig. 3e, f reveal the enrichment of oxygen vacancies to prove reduction in the PC2- region (that leads to enhanced migration of the SC/PC2- interface).

Moreover, we show that (*i*) an applied electric current can also induce AGG near the cathode (reduced region) in a simple polycrystalline specimen (Fig. 4a), and (*ii*) the grain growth of ZnO + 0.5 mol% Bi$_2$O$_3$ polycrystals can also be enhanced in reduced atmospheres (Ar-H$_2$ and Ar *vs.*

air) without an electric field (Fig. 4b-d). These observations further support that the oxygen reduction (instead of the electric field) is the direct reason that promotes grain growth.

**Reduction-induced GB disorder-order transitions.** To probe the atomic-level mechanism of the reduction-induced increases of GB mobilities, we further examine the underlying GB structures by AC STEM (Fig. 1). On the one hand, the slow-moving oxidized PC1+/SC interface is disordered (Fig. 1e, f), similar to the characteristic amorphous-like IGFs observed in a reference specimen with no electric field (Fig. 1a, b) that also exhibits similar low mobility (see Supplementary Note 2). On the other hand, the fast-moving and reduced PC2-/SC interface shows highly ordered structures (Fig. 1g, h). Furthermore, the GBs of the fast-growing abnormal grains in the reduced PC1- region are also ordered (Fig. 1c, d).

Additional examples of these four cases are given in Supplementary Figs. 9-12 to show the generality of the observations, and they are further discussed in Supplementary Note 5. Notably, various monolayer-, bilayer- and trilayer-like GB structures (presumably dependent on the specific crystallographic characters of the general GBs randomly selected from the specimen) have been observed in the reduced regions, but they are all ordered with high mobilities (in contrast to slow-moving amorphous-like GBs). Thus, we conclude that electrochemical reduction can generally induce disorder-to-order GB structural transitions with increased mobilities.

Fig. 5a, b further compares a pair of enlarged AC STEM high-angle annular dark-field (HAADF) images of GB structures at the oxidized PC1+/SC and the reduced SC/PC2- interfaces. Here, we also conduct image analyses to illustrate the layering and periodic orders (see Methods). While the amorphous-like GB (Fig. 5a) does show some partial orders (well-known for this type of IGFs[5,6]), it is more disordered and wider than the reduced GB that is highly ordered and bilayer-like in the case shown in Fig. 5b.

**Validation by first-principles calculations**. Density functional theory (DFT) calculations are critically compared with AC STEM images to further verify the reduction-induced GB disorder-to-order transition. Note that the bright contrasts at ZnO GBs in the HAADF images (see Fig. 1) are due to heavy Bi adsorbates (because of the $Z$ contrast). Such a strong Bi segregation has also been directly verified by energy dispersive X-ray spectroscopy (EDS), as shown in Supplementary Fig. 13. Based on the STEM images, we first construct an asymmetric GB to represent a general GB for DFT calculations, where one terminal plane was set to $(11\bar{2}0)$ to mimic both SC/PC interfaces observed in the STEM images (Fig. 5a, b). Next, the Bi coverage with a value of $\Gamma_{Bi}$ = ~11.7 nm$^{-2}$ was adopted to match prior experimental measurements of Bi segregation at ZnO GB at ~840-880°C [28] (see more details in Methods and Supplementary Note 6).

DFT structural optimization shows that the stoichiometric GB (representing oxidized conditions in experiments) exhibits a more disordered structure (Fig. 5e), while a reduced GB (after removing approximately one monolayer of oxygens) exhibits a more ordered bilayer-like structure (Fig. 5f; matching the STEM image in Fig. 5b). The bilayer-like Bi adsorption can be evident in the projected Bi distribution profile shown below Fig. 5f, with the interlayer distance of ~0.3 nm matching the STEM measurement (Fig. 5b). The simulated STEM images based on DFT-optimized structures (Fig. 5c, d) further verified disordered interfacial structure in stoichiometric GBs but ordered and bilayer-like Bi segregation structure in reduced GBs. Moreover, we calculate a structural disorder parameter ($\eta'$) for each atom in the DFT-relaxed structures and plot the projected disorder profiles ($\eta'(x)$) in Fig. 5g, h. For the disordered GB, the interfacial width is calculated to be ~0.8 nm based on the $\eta'(x)$ profile (Fig. 5g) and intensity profile in simulated STEM image (Fig. 5c), which agrees with STEM measured value of ~0.9 nm as shown in Fig. 5a (and ~0.7-0.9 nm for the different disordered GBs observed in this study). Further quantifications (by integrating the $\eta'(x)$ profiles) show that the stoichiometric GB has larger GB excess of disorder ($\Gamma_{Disorder} \approx 25$ nm$^{-2}$) than the reduced GB (~14 nm$^{-2}$); in other words, oxygen reduction makes the GB structure more ordered, which agrees with the experiments (Fig. 5a vs. 5b). We further simulated STEM images from DFT optimized GBs (Fig. 5c,d), which are consistent with the experiments (Fig. 5a,b) for the general disordered (Fig. 5a,c) vs. ordered (Fig. 5b,d) characters.

DFT calculations have also been conducted for GBs of different levels of Bi adsorption and oxygen reduction; see Supplementary Note 6 and Supplementary Figs. 14-15 for further details.

Furthermore, we calculate GB energies using the procedure described in Supplementary Note 7. Fig. 6a shows the DFT-computed GB energy difference $\Delta\gamma_{GB}$ ($\equiv \gamma_{GB}^{Reduced} - \gamma_{GB}^{Stoichiometric}$) as a function of oxygen chemical potential difference $\Delta\mu_O$ for both stoichiometric and reduced GBs. The crossover of two $\Delta\gamma_{GB}$ curves implies a GB phase-like or complexion transition from the disordered and stoichiometric GB (as shown by the orange solid line in Fig. 6a) to the ordered and reduced GB (as shown by the green solid line in Fig. 6a) with decreasing oxygen chemical potential, consistent with our experimental observations.

**A generalizable thermodynamic model.** To further discuss the physical origin of the reduction-induced GB disorder-to-order transition, we adopt a generalizable thermodynamic model following Tang *et al.*.[2] Here, the interfacial excess grand potential for a two-component GB in a diffuse-interface model is given by:[2]

$$\sigma^x = \int_{-\infty}^{+\infty}\left[\Delta f(\eta,c)+\frac{\kappa_\eta^2}{2}\cdot\left(\frac{d\eta}{dx}\right)^2+\frac{\kappa_c^2}{2}\cdot\left(\frac{dc}{dx}\right)^2+s\cdot g(\eta)\cdot\left|\frac{d\theta}{dx}\right|\right]dx, \tag{1}$$

where concentration $c(x)$, crystallinity $\eta(x)$ $(= 1 - \eta'(x))$, and crystallographic orientation $\theta(x)$, are functions of the spatial variable $x$, $\Delta f(c, \eta)$ is the homogenous free energy density referenced to the equilibrium bulk phases, and $\kappa_\eta$, $\kappa_c$, and $s$ are gradient energy coefficients. As derived in Supplementary Note 8, minimization of equation (1) leads to:

$$\sigma^x = s\cdot\Delta\theta\cdot\eta_{GB}^2 + \Delta F(\eta_{GB}), \tag{2}$$

where $\eta_{GB}$ is the order parameter at the center of the GB located at $x = 0$. Here, the first term $s\cdot\Delta\theta\cdot\eta_{GB}^2$ represents an energetic penalty to have a GB misorientation $\Delta\theta$, which can be lowered by GB disordering. The second term represents the total increased free energy due to the formation of a diffuse interface, which (initially) increases with GB disordering. Thus, the equilibrium level of GB disorder, $\eta'^{Equilibrium}_{GB} \equiv 1-\eta^{Equilibrium}_{GB}$ is determined by a tradeoff between these two terms. Specifically, equation (2) suggests that a smaller $s\cdot\Delta\theta$ will result in a more ordered GB at equilibrium state ($d(\sigma^x)/d(\eta^{Equilibrium}_{GB}) = 0$). This is further illustrated in Fig. 6b using a graphical construction method to solve equation (2) following Cahn's critical point wetting model,[32] as described in Supplementary Note 8.

Subsequently, we have developed a DFT-based method to estimate $s\cdot\Delta\theta$ to show that it (or $s$ since $\Delta\theta$ is a constant) can be decreased by about 2.4 time in the reduced GB in comparison with the stoichiometric GB; see Supplementary Note 8 and Supplementary Table 1 for detailed derivation and calculations. The smaller value of $s$ parameter ($s_2$) can result in a larger equilibrium GB order parameter $\eta^{Equilibrium}_{GB(2)}$ (Fig. 6b). Thus, the DFT calculations quantitively justify that oxygen reduction can induce a GB disorder-to-order transition.

**AIMD simulations of enhanced GB diffusivities.** We further perform large-scale *ab initio* molecular dynamics (AIMD) simulations to calculate and compare the GB diffusivities for the stoichiometric and reduced GBs. Fig. 6c shows the calculated GB diffusivities at 840 °C, which are increased markedly in the reduced and ordered GB in comparison with those in stoichiometric and disordered GB. For example, the Bi diffusivity increased by ~5×, Zn diffusivity increased by ~11×, and O diffusivity increased by ~2× in the oxygen reduced and ordered GB (Fig. 6c). Therefore, the AIMD simulations suggest the increased diffusion kinetics of the reduced and ordered GBs to explain the observed increased GB mobilities.

**Charge density maps and Bader charges.** The differential charge density maps obtained from DFT calculations further suggest that the increased diffusivity in the reduced GB can be attributed to the weaker charge transfer and chemical bonding (Supplementary Note 9 and Supplementary Fig. 18). Furthermore, we calculate average Bader charges to show that the effective charge on the Bi cations is decreased with the reduction (*i.e.*, one Bi atom losses ~1.4 *e* in the stoichiometric GB *vs.* ~0.69 *e* in the reduced GB; Supplementary Table 2).

**Discussion**

The above thermodynamic model and DFT and AIMD results can be understood intuitively. The presence of aliovalent $Bi^{3+}$ adsorbates (substituting $Zn^{2+}$ cations) in the stoichiometric GB likely promotes interfacial disordering. The oxygen reduction decreases the effective charge on Bi adsorbates to reduce interfacial disordering. These are supported by the calculated differential charge density maps and Bader charges discussed above.

Thus, we can envision the following mechanism. The aliovalent Bi adsorbates serve as charged "hot spots" to provide "pinning" effects at the stoichiometric GB with strong charge transfer (Supplementary Fig. 18) or chemical bonding. In contrast, the oxygen reduction can reduce the effective charge on Bi adsorbates (to weaken the bonding and alleviate "pinning" effects), thereby increasing the kinetics (diffusivities and mobilities) of the reduced GB. Finally, it is worthy to note that the observation of enhanced grain growth in reduced atmospheres (Fig. 4 and Supplementary Fig. 19) further supports our hypothesis that reduction can promote grain growth (even without an applied electric field/current); see Supplementary Note 10 for elaboration.

This work shows that $ZnO-Bi_2O_3$ can be used as a model system to uncover a fundamental mechanism of electrochemically induced GB transitions with significantly increased GB diffusivities, which subsequently result in enhanced and abnormal grain growth. These mechanisms have been further supported by controlled grain growth experiments and large-scale AIMD simulations. A generalized thermodynamic model associated with DFT calculations not only shed light on the physical origin of GB disorder-to-order transition, but also enable us to forecast other materials in future studies.

These findings have enriched our fundamental understandings of both electric field effects on microstructural evolution and the potentially transformative GB complexion (interfacial phase-like transition) theory via building a bridge between these two areas of great scientific importance and broad technological relevance.

Moreover, electrochemically induced GB transitions can exist in other systems and influence microstructural evolutions and various other properties, with potentially broad technological impacts on a variety of innovative materials processing technologies and electrochemical (or electronic) devices using electric fields and currents. This study also further suggests a new method to tailor the GB structure and properties electrochemically, as well as microstructures (*e.g.*, to intentionally produce graded and far-from-equilibrium microstructures).

## Methods

**Preparation of polycrystal/single crystal/polycrystal (PC/SC/PC) sandwich specimens.** 0.5 mol. % $Bi_2O_3$-doped ZnO powders were prepared by mixing ZnO (99.98% purity, ~18 nm, US Nanomaterials) with bismuth acetate (⩾99.99% purity, Sigma Aldrich). Mixed powders were ball-milled for 10 hours with a small amount of isopropyl alcohol. Powders were subsequently dried in an oven at 80 °C for 12 hours and annealed at 500 °C for 1 hour. ZnO ($11\bar{2}0$) single crystals with both sides polished were purchased from MTI Corporation (Richmond, California, USA). Dense $Bi_2O_3$-doped ZnO polycrystal/single crystal/polycrystal (PC/SC/PC) sandwich specimens were fabricated by spark plasma sintering (SPS) or field assisted sintering technology (FAST) at 780 °C for 5 minutes under a pressure of 50 MPa using a Thermal Technologies 3000 series SPS (Chatsworth, California, USA), and subsequently de-carbonized by annealing at 700 °C for 9 hours in air. After sintering, sandwich specimens reached >99% relative densities. Each sandwich specimen was ground to 5.0×5.0×~1.6 mm$^3$ cuboids with a 0.5-mm thick single crystal in between, which completely separated the two polycrystalline regions.

It is worth noting that the maximum solid solubility of $Bi_2O_3$ in the ZnO crystal is <0.06 mol% [28] so that most added $Bi_2O_3$ (0.5 mol% in the polycrystal) is present at grain boundaries (GBs) as the liquid-like nanoscale intergranular films (see, *e.g.*, Supplementary Fig. 9) or at triple-grain junction as a minor liquid phase at the annealing temperatures of 840-880 °C (above the ZnO-$Bi_2O_3$ eutectic temperature of 740 °C). Nevertheless, we adopt the term "$Bi_2O_3$-doped ZnO" that is commonly used in the ceramics field, albeit we acknowledge the majority of Bi is not doped into the ZnO crystal lattice.

**Annealing with an applied electric current.** Dense PC/SC/PC sandwich specimens were sputtered with platinum to form electrodes on both sides of two polycrystalline regions using a Denton Discover 18 Sputter (Moorestown, New Jersey, USA). An external DC electric current was applied on the specimen while annealing at a furnace temperature of 840 °C for 4 hours and maintaining a constant DC current density of $J = 6.4$ mA/mm$^2$. The specimen temperature, which was higher than the furnace temperature due to the Joule heating, was estimated to be ~865°C based on the electric power density and radiation heat transfer using a method described previously.[22] The electric field direction was perpendicular to the imbedded ZnO ($11\bar{2}0$) single crystal. More details of the experimental setup can be found in a prior publication (where the same setup was used for flash sintering experiments).33

The electric potentials and currents were recorded using a high-precision digital multimeter (Tektronix DMM 4050, Beaverton, Oregon, USA) during the experiments. The measured resistance *vs.* time curve for the PC/SC/PC sandwich can reach steady states after ~50 mins. See Supplementary Note 11 and Supplementary Fig. 20.

Following steady-state PC/SC/PC sample, all specimens were air quenched for characterization.

**Annealing experiments without an electric field/current**. To rule out the effects of the Joule heating with the applied electric current (that heated the specimen up by ~25 °C), the reference sandwich specimen was annealed at 880 °C for 4 hours using the exact same experimental setup but without any external electric field/current (*i.e.*, 40 °C higher in comparison with the furnace temperature of 840 °C for the specimen annealed with the applied electric current). As we noted in the prior section, the specimen temperature was estimated to be ~865 °C based on the Joule heating and radiation heat transfer[22] for the specimen annealed with the applied electric current. Thus, the annealing temperature of the reference specimen was (intentionally set to be) above the upper limit of the estimated specimen temperature of the case with an applied electric current. This ensures that any increased GB mobilities observed in the specimen with the applied electric current was not due to thermal effects (Joule heating).

The reference PC/SC/PC sandwich specimens were examined before and after the isothermal annealing (without an applied electric field/current), and the relevant SEM cross-sectional images are shown in Supplementary Figs. 6 and 7, respectively.

In addition, three dense polycrystalline specimens were prepared by using the same condition described above, and subsequently annealed at 880 °C for 4 hours without electric field/current in air, Ar, and Ar + 5% H$_2$, respectively, to investigate the effects of reducing atmospheres on grain growth. The results shown in Fig. 4b-d and Supplementary Fig. 19 further confirm that oxygen reduction can enhance grain growth.

After isothermal annealing, all specimens were air quenched for characterization.

**Characterization of microstructures and grain growth**. The densities of sandwich specimens were measured using the Archimedes method. Quenched specimens were characterized by using an ultra-high-resolution scanning electron microscope (Apreo SEM, FEI, Hillsboro, Oregon, USA) on the cross sections after grinding and polishing. Electron backscatter diffraction (EBSD) mapping was acquired by using an c-wave detector from Oxford Instruments (Concord, MA, USA).

The growth of the single crystal front was measured at 14 locations (×16 measurements per location) along each polycrystal-single crystal (PC/SC) interface for each case. At each location, the migration of the single crystal front (of the PC/SC interface) was averaged from 16 individual measurements with 5 μm intervals. The measured were conducted for both PC/SC interfaces in the specimens annealed with and without an applied electric field (for four cases all together). The results are plotted in Supplementary Fig. 2.

**Photoluminescence spectroscopy**. Spatially resolved photoluminescence spectroscopy in the wavelength range from 400 nm to 700 nm on the cross-sectional surface of the sandwich specimen was carried out on a confocal microscope (Leica SP5, Leica Microsystems, Wetzlar, Germany) equipped with a multiphoton system. The microscope spatial resolution is ~0.4 µm, and the penetration depth is ~50 nm.

**Aberration-corrected electron microscopy and energy dispersive X-ray spectroscopy.** Transmission electron microscopy (TEM) samples were prepared by using a dual-beam focused ion beam (FIB) / SEM system (Scios, FEI, Hillsboro, Oregon, USA) to lift out specimens of the selected GBs from the cross section based on the EBSD and SEM images.

Aberration-corrected scanning transmission electron microscopy characterization (AC STEM) of the GB structures was conducted by using a JEOL JEM-300CF STEM microscope (Akishima, Tokyo, Japan) operating at 300 kV. Both high-angle annular dark-field (HAADF) and bright field (BF) images were recorded.

Energy dispersive X-ray spectroscopy (EDS) was used in conjunction with STEM to confirm segregated regions (the GB complexions observed by STEM with bright contrast in HAADF imaging) are Bi-enriched (Supplementary Fig. 13).

**STEM image analyses to reveal order/disorder.** To examine the order and disorder (including the partial order in the amorphous-like GBs), we conducted two types of image analyses of the STEM images to reveal the layering *vs.* lateral orders, as follows.

First, we integrated the STEM HAADF intensities along the direction parallel to the GB to show the layering orders. Two examples of an amorphous-like GB (IGF) *vs.* an ordered bilayer are plotted as the yellow lines on the top of the STEM images in Fig. 5a, b.

Second, we conducted line-by-line fast Fourier transition (FFT) analysis to probe the lateral periodic orders. Here, we selected a rectangular frame (with the width being equal to the lattice spacing), moved the frame pixel by pixel, and conducted "line-by-line FFT analysis" of the crystalline order. Examples of an amorphous-like GB (IGF) *vs.* an ordered bilayer are plotted at the bottom panels in Fig. 5a, b.

**First-principles density functional theory (DFT) calculations.** The GBGenerator[34] code in Python Materials Genomics (pymatgen) library[35] was used to construct ZnO GB structure. The lattice parameters of the ZnO hexagonal structure ($a = 3.29$ Å and $c = 5.31$ Å) were taken from the Materials Project[36].

An asymmetric GB terminated by $(11\bar{2}0)$ plane (the two GBs shown in Fig. 5a, b) always requires a large simulation cell. Thus, we select the other GB plane to be the non-polar $(10\bar{1}0)$ with a rotate angle of ~53° along the [100] axis to construct a feasible model to mimic the GBs observed in the experiments (Fig. 5) within the limitation of the DFT cell size. Since lattice matching conditions cannot be achieved in three directions to apply periodic boundary conditions, a 15 Å thick layer of vacuum was added to isolate the interaction between two free surfaces created. The final simulation cell is triclinic with parameters: $a = 0.62$ nm, $b = 1.45$ nm, $c = 5.30$ nm, $\alpha = 104.69°$, $\beta = 78.69°$, and $\gamma = 74.70°$. This simulation cell contains 240 atoms in total, which is about the largest for effective DFT and AIMD simulations.

The first-principles DFT calculations were performed by using the Vienna *ab initio* Simulations Package (VASP) [37,38]. The Kohn-Sham equations were used to solve the projected-augmented wave (PAW) method [39,40] along with standard PAW potentials for the elements Zn, O, and Bi. The Perdew-Burke-Ernzerhof (PBE) [40] exchange-correlation functional was utilized to perform the structural optimization for the GB structure. The lattice parameters of ZnO were kept unchanged during the relaxation and only atomic positions were subjected to relaxation. All atoms were fully relaxed until the Hellmann-Feynman forces were smaller than 0.02 eV/Å. The Brillouin-zone integrations were sampled on a $\Gamma$-centered 4×2×1 *k*-point grids. The kinetic energy cutoff for plane waves was set to 400 eV. The convergence criterion for electronic self-consistency was adopted to $10^{-4}$ eV.

**STEM simulation.** STEM HAADF images were simulated by using the QSTEM program. The DFT-optimized GB structures of stoichiometric and reduced Bi-doped ZnO GBs were adopted for imaging simulation. We adopted the electron voltage with a value of 300 kV (same as experiments) for all STEM simulations. The scattering semi-angle for HAADF imaging was set to 60 mrad, the convergence angle was adopted to 20 mrad, and the spherical aberration coefficient was set to 0.5 μm.

**Quantification of GB order/disorder from DFT simulations.** A bond-orientational order parameter $\eta$ was calculated for each atom[41]. Subsequently, we defined a dimensionless disorder parameter ($\eta' \equiv 1 - \eta$; $\eta' = 1$ for an atom in a liquid and $\eta' = 0$ for an atom in a perfect crystal).

The 1D distribution of disorder parameter $\eta'(x)$ was obtained by averaging the $\eta'$ values of atoms in the directions parallel to the GB plane. Two examples of disorder parameter profiles $\eta'(x)$ for the amorphous-like GB (IGF) *vs.* the ordered bilayer are shown in Fig. 5g, h. Subsequently, GB excess disorder was quantified by integrating the disorder profile $\eta'(x)$. More details can be found in prior publications where similar analyses were performed for atomistic simulations using empirical potentials[9,42,43] (*vs.* extending the definitions and methodologies to DFT relaxed GB structures here in this study).

**Quantification of atomic density profiles from DFT simulations.** We used a coarse-grained method to compute 1D atomic density profiles from DFT-relaxed GB structures to compare with the STEM results. In this procedure[9], a Gaussian function was assigned to each atom and the overall density distribution function can be computed by summing all Gaussian functions. Two examples of Bi profiles for the amorphous-like GB *vs.* the ordered bilayer are shown in Fig. 5e, f.

***Ab initio* molecular dynamics (AIMD) simulations of GB diffusivities.** Based on the optimized GB structures, we performed *ab initio* molecular dynamics (AIMD) simulations under the *NVT* ensemble with a Nose-Hoover thermostat[44,45] to obtain GB diffusivities. The temperature was set to 1123 K (850 °C), close to the experimental annealing condition. The overall simulation time was set to 1500 fs with a time step of 1 fs. Although the overall simulation time is relatively short (due to the limitation of a very large simulation cell), we believe the GB structures can achieve equilibrium based on monitoring the potential energy *vs.* time. The *k*-point grids were adopted to 1×1×1 ($\Gamma$ point only). To avoid the effects of vacuum, we fixed 2~3 monolayer atoms near the free surfaces and only allowed other atoms to move. The atoms' trajectories during AIMD simulation were used to calculate the mean square displacement (MSD) over time (*t*). Finally, the GB diffusivities for Zn, O and Bi, respectively, were obtained by linearly fitting the corresponding MSD *vs. t* curves for both stoichiometric and reduced GBs for comparison (Fig. 6c).

**Data availability**

All data are available from the corresponding authors on reasonable request.

**Acknowledgments:** This work is supported by the Aerospace Materials for Extreme Environments program of the U.S. Air Force Office of Scientific Research (AFOSR) under the grants nos. FA9550-14-1-0174 (2024-2019) and FA9550-19-1-0327 (2019-2022).

**Author contributions:** J. L. conceived the idea and supervised the work. J. N. conducted most experiments and Q. Y assisted some experiments and data analysis. C.H. performed DFT calculations and AIMD simulations. All authors drafted, reviewed, and revised the manuscript.

**Competing interests:** The authors declare no conflict of interests.

**Supplementary Information:**
Supplementary Notes 1-11
Supplementary Tables 1-2
Supplementary Figs. 1-20
Supplementary References

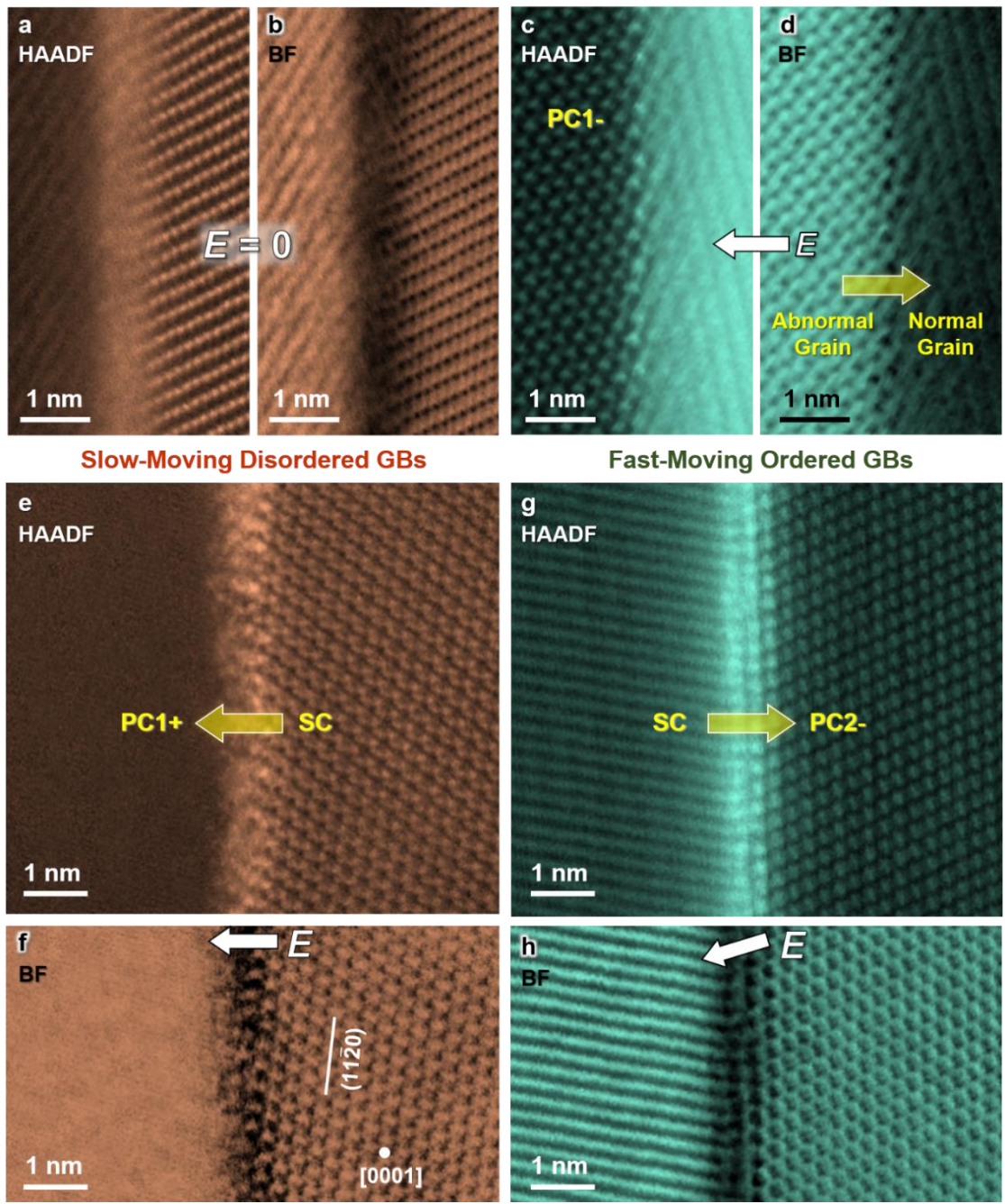

**Fig. 1 AC STEM high-angle annular dark-field (HAADF) and bright-field (BF) images of representative slow-moving disordered *vs.* fast-moving ordered GBs. a**, **b**, A nanoscale amorphous-like IGF (*a.k.a.* disordered GB) observed in a reference specimen annealed without an electric field, which is characteristic of all general GBs in $Bi_2O_3$-saturated ZnO.[29] STEM images of (**c**, **d**) an ordered GB of an abnormal grain in the electrochemically reduced PC1- region, as well as (**e**, **f**) a slow-moving disordered GB at the oxidized PC1+/SC interface *vs.* (**g**, **h**) a fast-moving ordered GB at the reduced SC/PC2- interface in a PC1/SC/PC2 sandwich specimen annealed with an applied electric current (as schematically shown in Fig. 2a). STEM images for additional examples can be found in Supplementary Figs. 9-12, showing the generality of the observations of slow-moving disordered GBs *vs.* fast-moving ordered GBs.

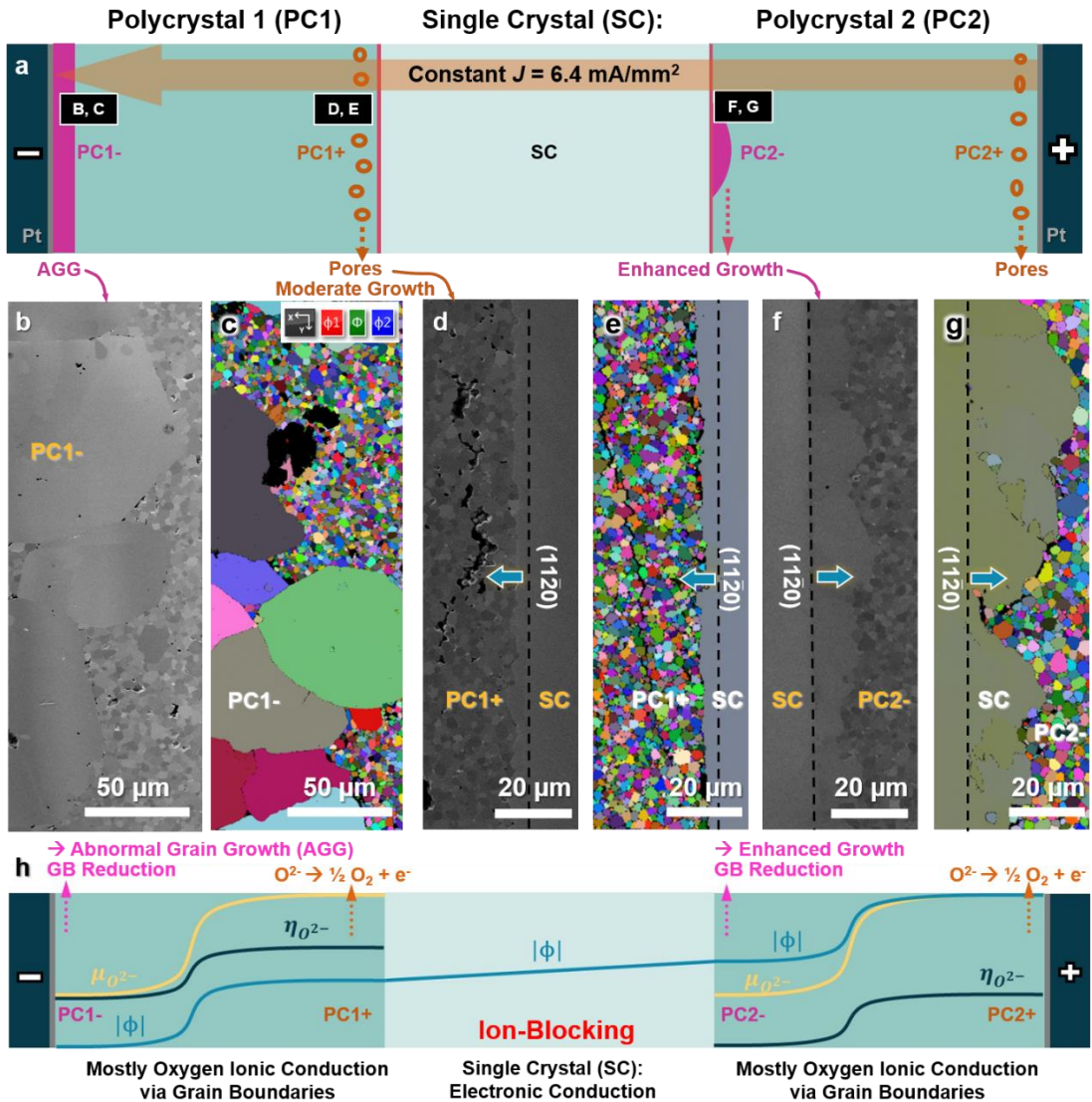

**Fig. 2 Microstructure of a Bi$_2$O$_3$-doped ZnO sandwich specimen isothermally annealed under a constant applied current. a**, A schematic illustration of this PC1/SC/PC2 sandwich specimen. Enhanced or abnormal grain growth takes place in the reduced PC1- and PC2- regions. Cross-sectional scanning electron microscopy (SEM) micrographs and electron backscatter diffraction (EBSD) Euler maps of (**b**, **c**) abnormal grain growth (AGG) near the cathode in the reduced PC1- region, as well as (**d**, **e**) moderate migration of the oxidized PC1+/SC interface *vs.* (**f**, **g**) enhanced migration of the reduced SC/PC2- interface. Dashed lines indicate the original positions of the SC/PC interfaces. See Supplementary Fig. 2 for quantitative measurements of migration distances of both PC1+/SC and SC/PC2- interfaces (in comparison with those in a reference sandwich specimen without an applied electric field) and Supplementary Figs. 3-5 for addition EBSD maps. **h**, Schematic profiles of electric ($\phi$), chemical ($\mu_{O^{2-}}$), and electrochemical ($\eta_{O^{2-}}$) potentials *vs.* locations. See Supplementary Fig. 1 for further detail.

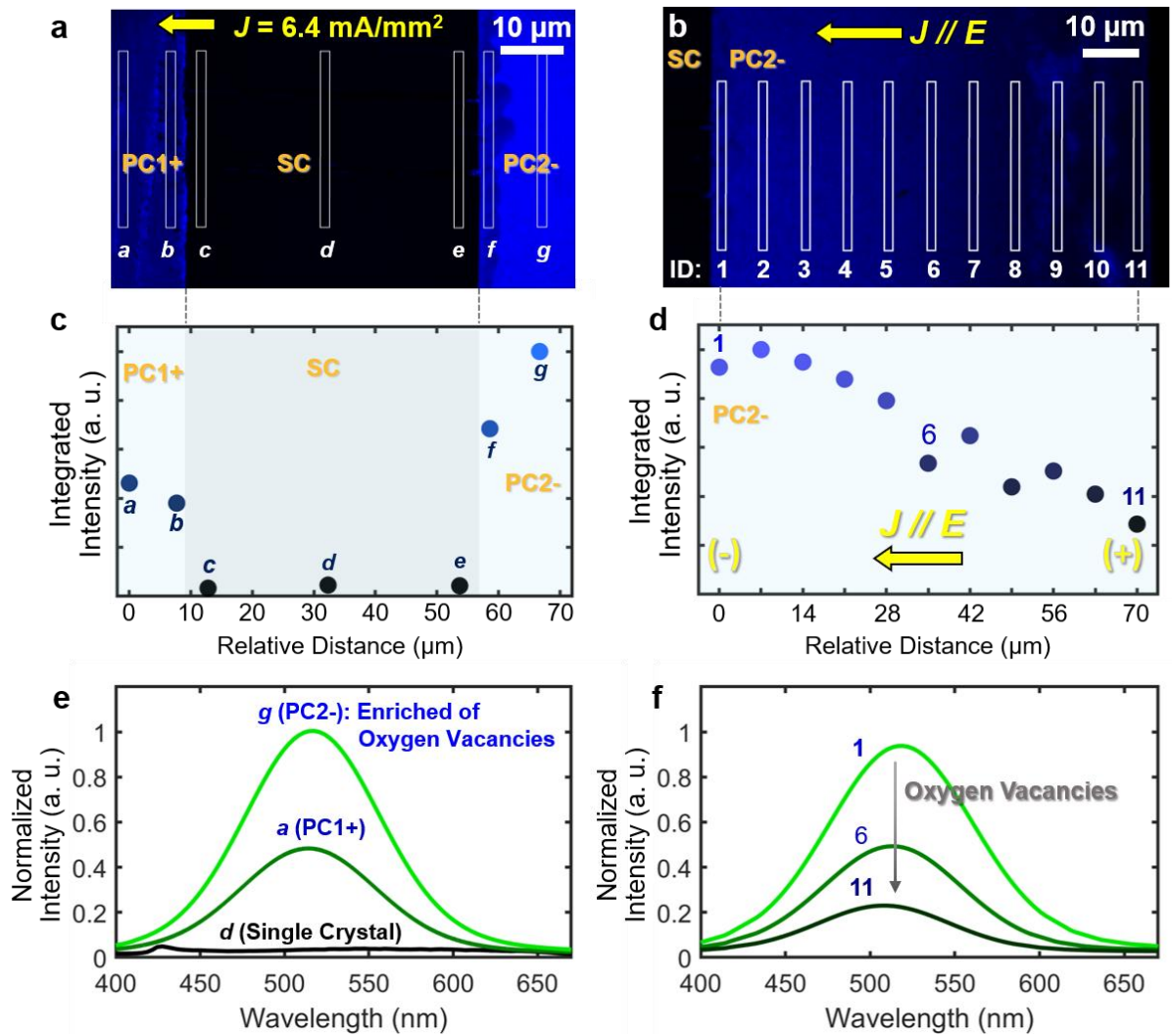

**Fig. 3 Photoluminescence spectroscopy of the sandwich specimen annealed with a constant applied electric current, suggesting the enrichment of oxygen vacancies in the reduced PC2- region. a**, **b**, Maps of the photoluminescence intensity at the 526 nm wavelength of the cross-sectional PC1/SC/PC2 sandwich specimen. **c**, **d**, Integrated photoluminescence intensities for the combined photoluminescence peak at ~400-700 nm at different locations, representing all defects at GBs. Photoluminescence spectra are collected at Locations *a-g* labeled in Panel (a) and Locations 1-11 labeled in Panel (b), respectively. **e**, **f**, the normalized photoluminescence intensity *vs.* wavelength curves of decomposed green-emission band (representing the oxygen vacancy concentration) at selected locations. See Supplementary Note 4 and Supplementary Fig. 8 for further detail. In Panel (c-f), a.u. = arbitrary unit.

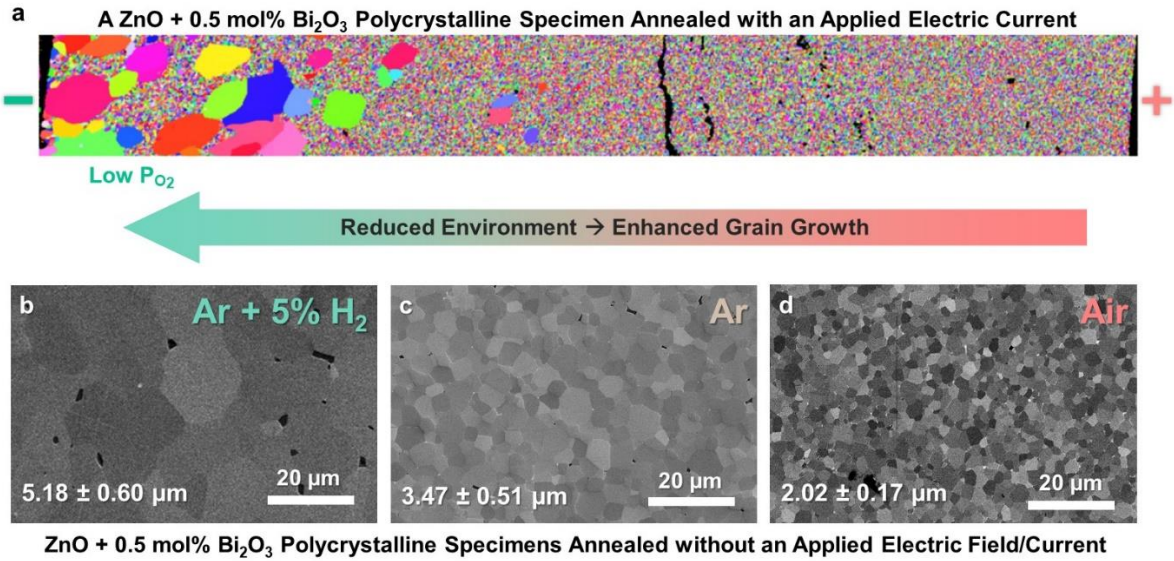

**Fig. 4 Enhanced grain growth in reduced environments for ZnO + 0.5 mol% $Bi_2O_3$ polycrystalline specimens. a**, EBSD map of a ZnO + 0.5 mol% $Bi_2O_3$ polycrystalline specimen (without a single-crystal section) annealed with an applied current, showing similar abnormal grain growth in the reduced region near the cathode (negative electrode). Cross-sectional SEM micrographs of polycrystalline specimens quenched from 880°C after annealing for 4 hours in **b**, Ar + 5% $H_2$ **c**, Ar, and **d**, air. Additional low-magnification images are shown in Supplementary Fig. 19. These observations further support that the oxygen reduction caused enhanced grain growth.

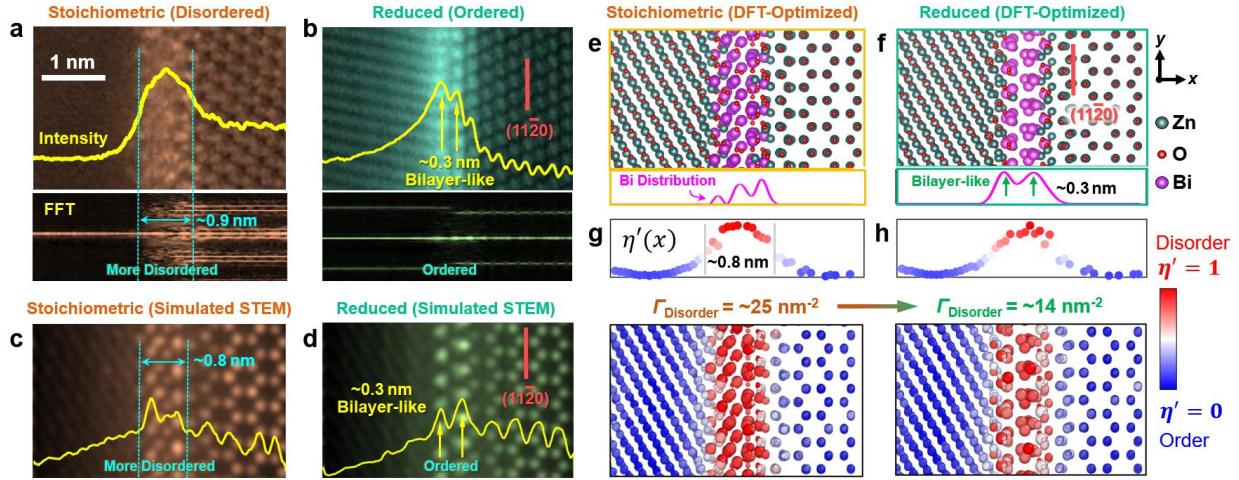

**Fig. 5 Comparison of experiments and DFT simulations of stoichiometric (disordered) *vs.* reduced (ordered) GB structures and AIMD simulated GB diffusivities.** Expanded experimental STEM HAADF images for (**a**) a disordered and stoichiometric GB *vs.* (**b**) an ordered (bilayer-like) and reduced GB; here, we plot the averaged HAADF intensity and line-by-line FFT patterns to illustrate the layering and periodic orders. Simulated STEM HAADF images for (**c**) stoichiometric GB *vs.* (**d**) reduced GB, respectively, based on DFT-optimized structures of (**e**) the stoichiometric *vs.* (**f**) reduced GBs, where the projected Bi density profiles are plotted beneath. The calculated disorder parameters for all atoms for DFT-optimized (**g**) stoichiometric *vs.* (**h**) reduced GB structures. The projected disorder parameter profiles ($\eta'(x)$) are shown above. The GB excess of disorder (computed by integrating $\eta'(x)$) decreases from $\Gamma_{\text{Disorder}}$ = ~25 nm$^{-2}$ for the stoichiometric GB to $\Gamma_{\text{Disorder}}$ = ~14 nm$^{-2}$ for the reduced GB, thereby showing that the oxygen reduction induces GB ordering.

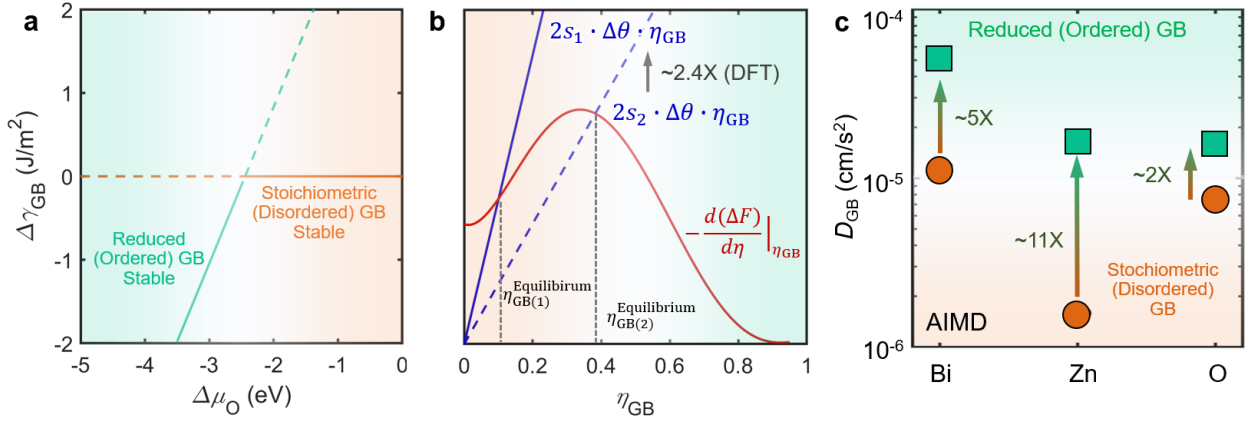

**Fig. 6 DFT-calculated interfacial energetical diagram, schematic diagram of GB disorder-to-order transition based on a generalized thermodynamic model, and *ab initio* molecular dynamics (AIMD)-simulated GB diffusivities. a**, Computed GB energy difference $\Delta\gamma_{GB}$ ($\equiv \gamma_{GB}^{Reduced} - \gamma_{GB}^{Stoichiometric}$) *vs.* oxygen chemical potential difference $\Delta\mu_O$ ($\equiv \mu_O - \frac{1}{2}E_{O_2}$), showing a transition from the stoichiometric (disordered) GB to the reduced (ordered) GB with decreasing oxygen chemical potential, consistent with experiments. **b**, Schematic illustration of a graphical construction method to solve the equation (2). The intersection of the red and blue lines indicates an equilibrium order parameter, *e.g.*, $\eta_{GB(1)}^{Equilibrium}$. DFT calculations showed that the parameter *s* for the stoichiometric GB ($s_1$) is about ~2.4× of reduced GB ($s_2$), suggesting that oxygen reduction can lead to larger equilibrium order parameter $\eta_{GB(2)}^{Equilibrium}$ and thus explain the GB disorder-to-order transition. **c**, GB diffusivities calculated by AIMD simulations. The GB diffusivities in the reduced (ordered) GB are markedly increased in comparison with those in the stoichiometric (disordered) GB.

# Supplementary Information

# Discovery of Electrochemically Induced Grain Boundary Transitions

Jiuyuan Nie [1,#], Chongze Hu [1,2,#], Qizhang Yan [1], and Jian Luo [1,2,*]

[1] Department of Nanoengineering; [2] Program of Materials Science and Engineering

University of California, San Diego

La Jolla, California 92093, U.S.A.

[#] These authors contributed equally.

[*] Corresponding author. E-mail address: jluo@alum.mit.edu (J. Luo).

# Table of Contents

# Supplementary Note 1:
## Discussion of electric effects on microstructural evolution and related innovative processing technologies

Innovative Sintering Technologies:

Electric fields and currents are used in flash sintering [1,2], electro-sintering [3,4], and field-assisted sintering technology (FAST; commonly known as "spark plasma sintering" or "SPS") [5,6].

First, in the "flash sintering" pioneered by Raj and co-workers [2], ultrafast densification in seconds at reduced furnace temperatures is enabled via directly flowing electric currents through specimens [1,2,7,8]. While recent studies demonstrated that the flash generally initiates as a thermal runaway [9-11] and ultrafast densification stems from the ultrahigh heating rates [12,13], electric fields and currents can significantly influence microstructural evolutions [9,14-21] and induce other unusual phenomena [17,22-25]. Notably, asymmetrical grain growth has been widely observed in flash sintered ZnO [9], ZrO$_2$ [26], 3YSZ [27], MgAl$_2$O$_4$ [28], and UO$_2$ [29], among others [1,17,19], but their underlying mechanisms are largely unknown. Possible roles of grain boundary (GB) complexions on flash sintering have been discussed [12,13,30,31] and explored [7,32], but no in-depth study has been conducted to reveal how they may influence grain growth.

Second, in the "electro-sintering" discovered by Chen and co-workers, an large electric current was used to enhance densification of 8 mol% Y$_2$O$_3$-stabilized ZrO$_2$ (8YSZ) via ionomigration of pores at a low specimen temperature [3,4], which can have drastic impacts on microstructural evolution (to be elaborated subsequently) at the same time [33]. Thus, understanding how electric fields and currents influence the microstructural evolution can not only improve these novel sintering technologies, but also provide new possibilities to control or even tailor microstructures.

Third, electric fields are known to affect the microstructural evolution in the widely used FAST/SPS [19] even if a large portion of the electric current may flow through the graphite tooling surrounding the specimen in the conventional setting (albeit that the electric current can be forced into the specimen via the "flash SPS" setting [34-36]).

A similar effect may also be expected for the most recently reported ultrafast high-temperature sintering (UHS) that can densify ceramics in ~10 seconds [37], where the current is mostly running through the thin carbon heaters around the specimen but the electric field may still affect the microstructural evolution near the specimen surface.

Electrochemical Devices for Energy Storage and Conversion:

Electric fields and currents are present in solid electrolytes used in solid oxide fuel cells [38-40] and solid-state batteries [41], as well as various other electrochemical or electronic devices that use electric fields and currents, where they can cause unexpected (usually undesirable) changes in microstructures. The underlying mechanisms are controversial.

Specifically, fascinating yet often elusive observations of the electric effects of grain growth are briefly discussed using three model systems as exemplars, as follows.

YSZ ($Y_2O_3$-Stabilized $ZrO_2$):

Earlier studies already revealed interesting and intriguing observations of the electric field effects on grain growth of YSZ. For example, Conrad *et al.* showed that a relatively weak applied DC or AC field could inhibit grain growth of 3YSZ and attributed it to the interactions of the applied electric fields with space charges [42-46]. In contrast, Chen and colleagues demonstrated that a large applied electric current of ~50 A/cm$^2$ could enhance the GB mobility by >10 times in the cathode (negative electrode) side discontinuously in 8YSZ [33]. A series of follow-up studies further confirmed the generality of, and investigated, this cathode-side enhanced grain growth phenomenon in several fluorite-type oxides (including 3YSZ) [38,47-49]. Here, Chen and coworkers attributed the enhanced grain growth in the cathode side to the electrochemically driven reduction that lowers the GB migration barriers (from the bulk defects and bulk diffusion point of view, instead of possible GB transitions, which have not yet been examined), and developed a bulk defect chemistry based model [38,47-49]. They further showed that an "oxygen potential transition" in the bulk can induce a phase-like transition behaviors in grain growth [38]. It is yet unknown whether applied electric fields and currents can also change the GB structures to influence the microstructural evolution in YSZ.

Pervoskite $SrTiO_3$:

Rheinheimer *et al.* showed that a weak electric field (with no current, via using "blocking" electrodes) can promote grain growth of $SrTiO_3$ near the negative electrode, which was hypothesized to be resulted from the increased concentrations of oxygen vacancies [50]. However, the exact atomic-level mechanism of how oxygen vacancies can enhance grain growth is unclear. Separate bicrystal experiments by Hughes and van Benthem suggested that an applied electrostatic field (again with no current) could change the structure of (100) symmetric twist GBs in $SrTiO_3$, which was explained from asymmetric "ordering of the oxygen sublattice" in the bicrystals of a special geometric configuration [51,52]. It is unknown whether such structural transitions can also exist at general GBs. In flash-sintered $SrTiO_3$ (with a current), Rheinheimer *et al.* further revealed preferential Ti segregation at general GBs near the positive electrode [53]. However, many open scientific questions remain; a complete understanding of how and why an applied electric field/current can alter the GB structure and/or chemistry and how they subsequently change the grain growth behaviors, as well as the exact underlying mechanisms, have not been established.

ZnO and $Bi_2O_3$-doped ZnO:

For flash sintering of undoped ZnO in air, our prior study revealed an anode-side abnormal grain growth [9], in contrast to the cathode-side enhanced grain growth observed in YSZ [33,47-49] and $SrTiO_3$ [50]. This was explained from a hypothesized GB oxidation transition [7]. This hypothesis was supported indirectly by the suppression of this anode-side abnormal grain growth during flash sintering of pure ZnO in reduced atmospheres [7,54]. However, we could not directly confirm and characterize this hypothesized GB oxidation transition in undoped ZnO (in part because it is difficult, if not impossible, to quench the undoped ZnO specimens to examine the atomic-level GB structures).

In this study, we used $Bi_2O_3$-doped ZnO (albeit the solid solubility of $Bi_2O_3$ in the ZnO crystal is <0.06 mol% [55] so that most added $Bi_2O_3$ is present at GBs or as a secondary phase) to explicitly show, for the first time to our knowledge, that an applied electric current can induce a GB disorder-order transition that can be well quenched for direct characterization by aberration-corrected scanning transmission electron microscopy (AC STEM). Moreover, first-principles calculations confirmed this new mechanism of electrochemical reduction induced transition and *ab initio* molecular dynamics (AIMD) simulations further showed enhanced GB diffusion in reduced and ordered GBs, which fully explained the observed enhanced and abnormal grain growth. Interestingly, enhanced grain growth occurred in the reduced region in $Bi_2O_3$-doped ZnO near the cathode, in contrast that in the oxidized region near the anode for undoped ZnO [9]. Here, the $Bi_2O_3$ segregation at GBs enabled us to better quench the interfacial structures for detailed characterization and subsequent modeling, thereby establishing a complete link of the underlying mechanisms of how an applied electric field induce a GB structural transition to lead to enhanced grain growth, for the first time in this study. Moreover, a generalizable thermodynamic model has been established for the reduction induced GB disorder-order transitions.

# Supplementary Note 2:

## Microstructural evolution of the sandwich specimens (with and without an applied electric current)

The Polycrystal 1/Single Crystal/Polycrystal 2 (PC1/SC/PC2) specimen used in the grain growth experiment with a constant applied electric current is schematically shown in Supplementary Fig. 1 below.

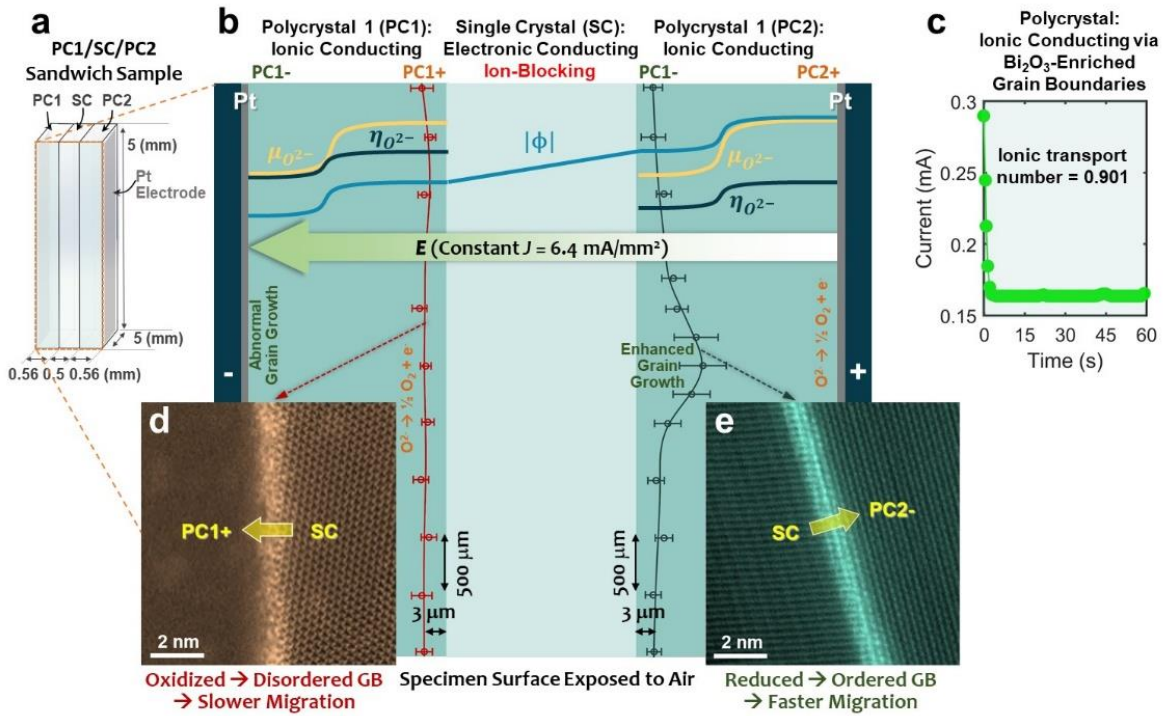

**Supplementary Fig. 1. The schematic illustration of the sandwich grain growth experiment with a constant applied electric current. a**, Schematic drawing of the sandwich sample. **b**, Proposed profiles of electric ($\phi$), chemical ($\mu_{O^{2-}}$), and electrochemical ($\eta_{O^{2-}}$) potentials *vs.* locations. The measured migration distances at PC1+/SC and SC/PC2- regions are plotted. Noting that the aspect ratios are not drawn to the scale. Grain boundaries (GBs) in the two polycrystals are mostly ion-conducting via the $Bi_2O_3$-enriched liquid-like intergranular films (IGFs), while the ZnO single crystal is electron-conducting but ion-blocking. **c**, The ionic transport number measurement of a ZnO + 0.5 mol% $Bi_2O_3$ polycrystalline specimen at 840 °C. The measured high ionic transport number of ~0.9 indicates that the specimen is mostly ion-conducting (via the $Bi_2O_3$-enriched liquid-like IGFs). Consequently, PC1- and PC2- regions must be reduced, while PC1+ and PC2+ regions must be oxidized. See Supplementary Note 3 for detailed discussion. Representative STEM HAADF images of (**d**) a slow-moving disordered GB at the oxidized PC1+/SC interface and (**e**) a fast-moving ordered GB at the reduced SC/PC2- interface.

The key observations of the "unusual" microstructural evolutions in the PC1/SC/PC2 specimen induced by an applied electric current are summarized as follows:

(1) The PC1+/SC interface, which was oxidized (as indicated by the generation of pores presumably due to the oxidation reaction that produced $O_2$ gas: $O^{2-} \rightarrow \frac{1}{2}O_2 + e^-$), only had moderate migration of ~3 μm (Fig. 2d, e and Supplementary Fig. 4).

(2) In contrast, abruptly enhanced migration of the SC/PC2- interface was observed (Fig. 2f, g

and Supplementary Fig. 3) in the reduced middle section of the SC/PC2- interface (to a maximum of ~10 μm at the center, which significantly increased GB mobility in comparison with both the oxidized PC1+/SC interface and the two PC/SC interfaces in the reference specimen without an applied electric field/current, as elaborated subsequently).

(3) Abnormal grain growth took place in the reduced PC1- region near the negative electrode (cathode) that resulted in extremely large grains, as clearly evident in the SEM images and corresponding EBSD maps in Fig. 2b, c and Supplementary Fig. 5.

To better quantify the grain growth, two reference specimens were fabricated and characterized (both with symmetric grain growth without an applied electric field/current), including:

*(i)* An as-sintered PC/SC/PC sandwich specimen (prior to annealing at 880 °C), where the averaged migration distances at two PC/SC interfaces were measured to be 2.81 ± 0.94 μm and 2.72 ± 0.88 μm, respectively (Supplementary Fig. 6).

*(ii)* Another reference PC/SC/PC sandwich specimen annealed at 880 °C for 4 hours without an external electric field/current, where the averaged migration distances at two PC/SC interfaces were measured to be 3.63 ± 0.91 μm and 3.60 ± 0.79 μm, respectively (Supplementary Fig. 7).

Supplementary Fig. 2 shows a quantitative comparison of the PC1/SC/PC2 specimen annealed with an applied electric current at a furnace temperature of 840 °C (with an estimated specimen temperature of ~865 °C considering the moderate Joule heating) for 4 hours and the reference specimen (discussed in Case (*ii*) above) that was annealed without an applied electric field/current at 880 °C (intentionally set to be slightly higher that the upper limit of specimen temperature in the case with an applied electric current) for 4 hours. A critical comparison of quantitative measurements of the migration distances *vs.* lateral locations of four SC/PC interfaces is shown in Supplementary Fig. 2 and discussed as follows:

(A) The migration distances of the oxidized SC/PC1- interface of the specimen annealed with an applied electric current were measured to be 3.04 ± 0.82 μm (*i.e.*, slightly lower than the ~3.6 μm in the reference specimen without an applied electric current).

(B) The migration distances of the PC1+/SC interface of the specimen annealed with an applied electric current were significantly increased in the middle (mostly reduced) section of the specimen to a maximum of ~10 μm growth at the center. The enhanced growth occurred abruptly in the middle (~10-15% of the specimen width, forming a ~500 μm wide protrusion), more reduced region; it decayed to normal growth distances of ~3-4 μm near the (less reduced) free surfaces that were exposed to air, as shown in Supplementary Fig. 2

In summary, the increased GB mobilities occurred in the reduced PC1- and PC2- regions, while the grain growth was moderate in the oxidized region, on a par with that in the reference specimen without an applied electric field/current. Quantitative measurements and a careful comparison have clearly demonstrated that an applied electric current can abruptly increase GB mobility in the electrochemically reduced regions.

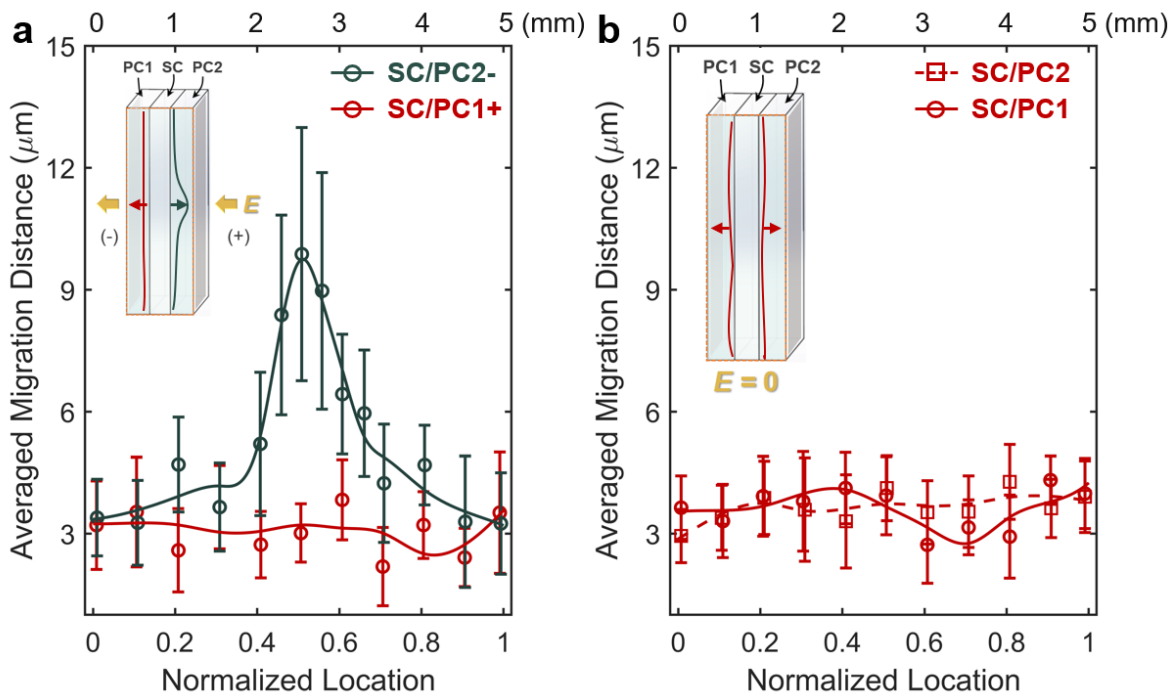

**Supplementary Fig. 2. Measured averaged migration distances of the single-crystal fronts *vs.* normalized location (to a 5-mm total length) of Bi$_2$O$_3$-doped ZnO PC1/SC/PC2 sandwich specimens annealed with and without an applied electric current.** The migration distances the single crystal fronts at both sides were measured from cross-sectional SEM images at the two SC/PC interfaces in (**a**) one sandwich specimen annealed at a furnace temperature of 840 °C with a constant current density of $J$ = 6.4 mA/mm$^2$ (with an estimated specimen temperature ~865 °C considering the moderate Joule heating) and (**b**) another reference sandwich specimen annealed at 880 °C without electric field/current ($E$ = 0). Both specimens were annealed for 4 hours and subsequently quenched. Each data point was averaged from 16 individual measurements with 5 µm intervals (with the standard deviation as the error bar). Abruptly enhanced growth was observed in the intermediate (*i.e.*, more isolated and reduced) section of the SC/PC2- interface in the specimen annealed with an applied electric current. It is noted that the annealing temperature (880 °C) of the reference specimen was intentionally set to be slightly higher than the estimated specimen temperature (of ~865 °C) for the case with the applied electric current to unequivocally demonstrate that the observed enhanced migration of the SC/PC2- interface is not a temperature (Joule heating) effect.

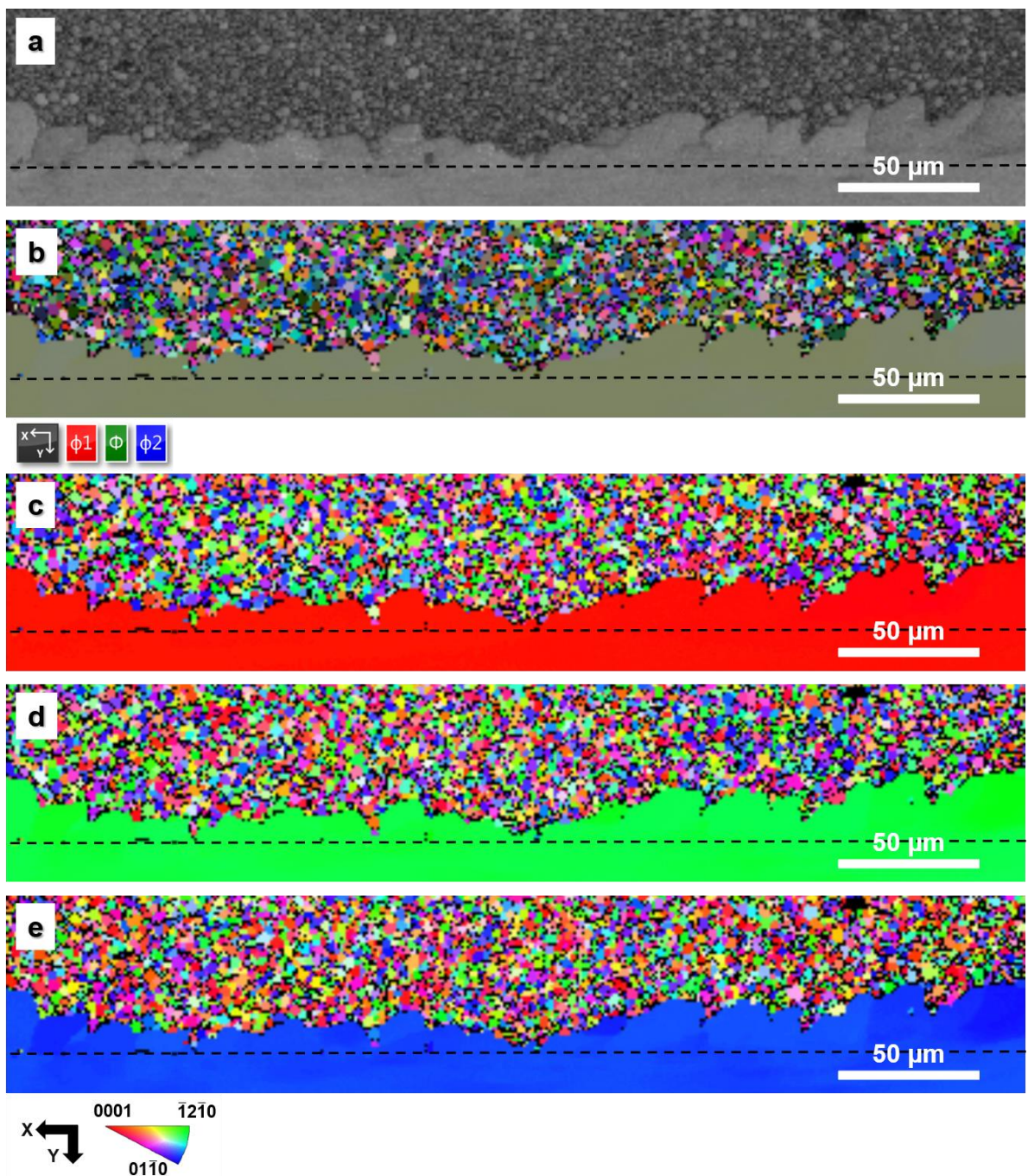

**Supplementary Fig. 3. EBSD mapping of the reduced SC/PC2- region with enhanced growth in the $Bi_2O_3$-doped ZnO sandwich specimen annealed with a constant electric current density of $J$ = 6.4 mA/mm$^2$. a**, Band contrast of the EBSD map, which shows Kikuchi bands. **b**, Euler maps of the SC/PC2- interface, showing the enhanced growth along the original single crystal orientation. **c-e**, The X, Y, and Z direction inverse pole figure (IPF) orientation maps. The color scheme and reference orientations are shown in the legends.

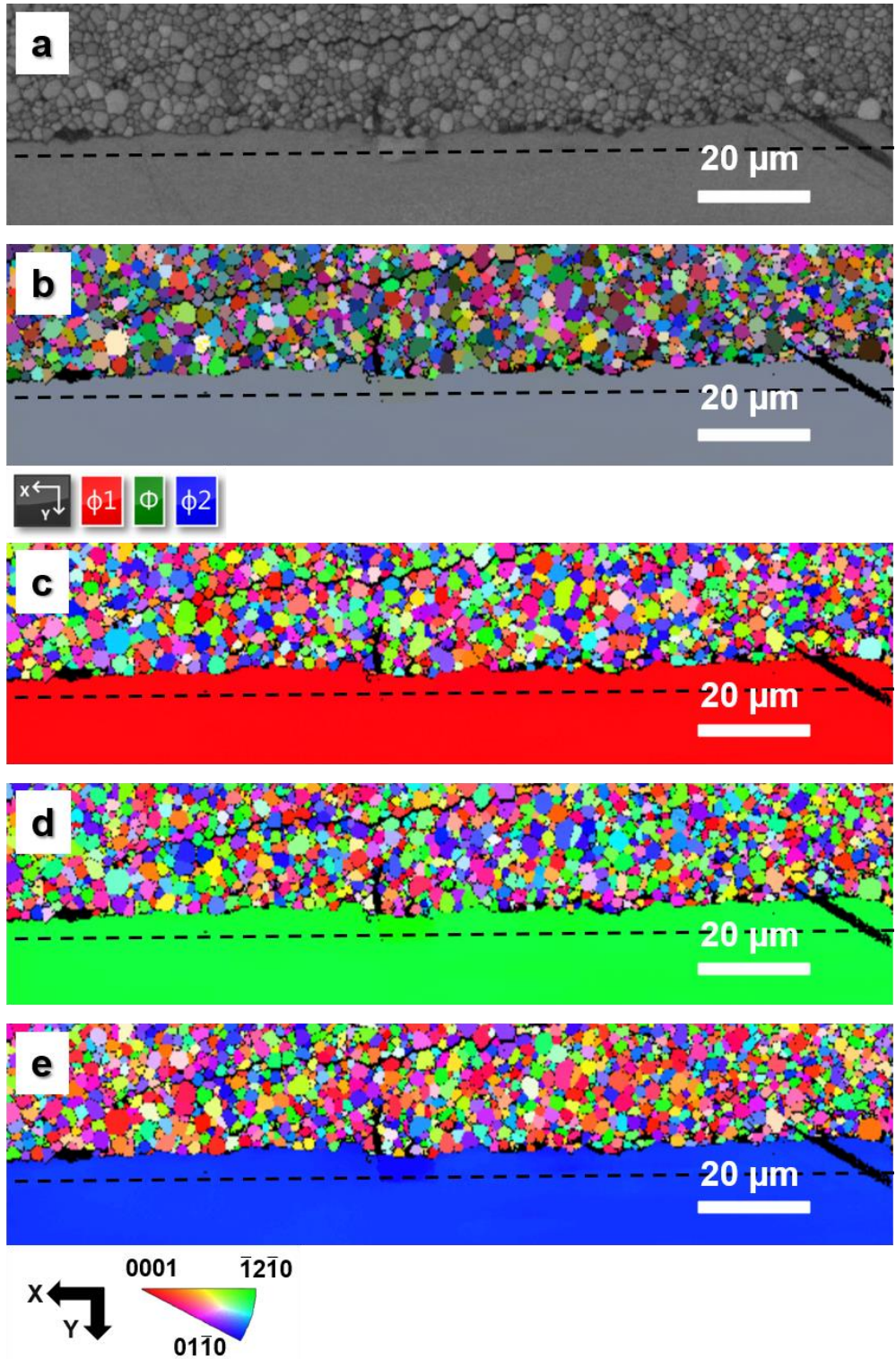

**Supplementary Fig. 4. EBSD mapping of the oxidized SC/PC1+ region without enhanced growth in the $Bi_2O_3$-doped ZnO sandwich specimen annealed with a constant electric current density of $J$ = 6.4 mA/mm$^2$. a**, Band contrast of the EBSD map, which shows Kikuchi bands. **b**, Euler maps of the SC/PC1+ interface, showing the growth along the original single crystal orientation. **c-e**, The X, Y, and Z direction inverse pole figure orientation maps. The color scheme and reference orientations are shown in the legends.

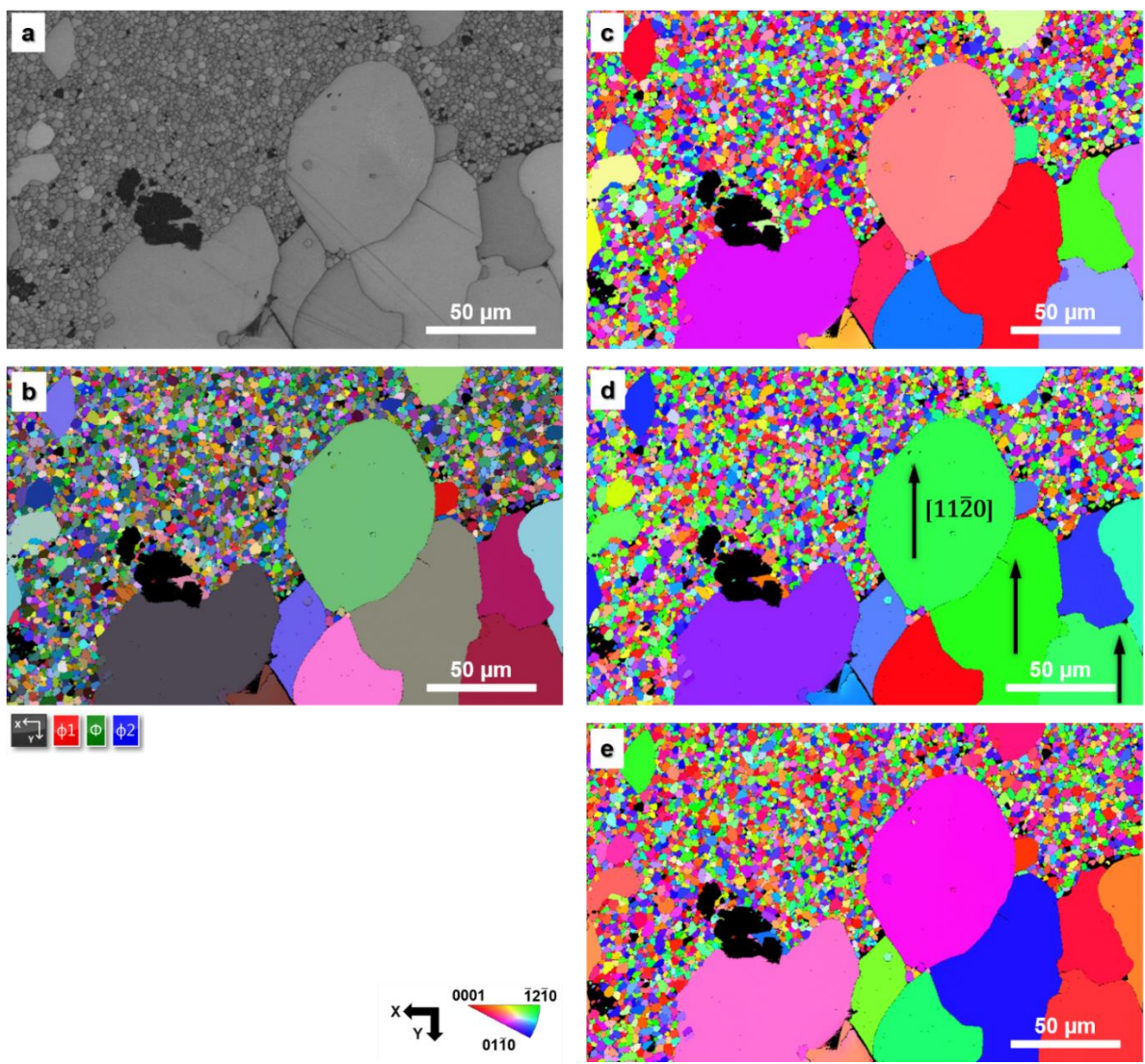

**Supplementary Fig. 5. EBSD mapping of the abnormal grains near the cathode side (in the electrochemically reduced PC1- region) in the Bi$_2$O$_3$-doped ZnO sandwich specimen annealed with a constant electric current density of $J$ = 6.4 mA/mm$^2$. a**, Band contrast of the EBSD map, which shows Kikuchi bands. **b**, Euler maps of the PC1- region, showing the abnormal grain growth. **c-e**, The X, Y, and Z direction inverse pole figure (IPF) orientation maps. The color scheme and reference orientations are shown in the legends. A preferential crystallographic orientation, [11$\bar{2}$0], is identified in Panel (d) Y-IPF.

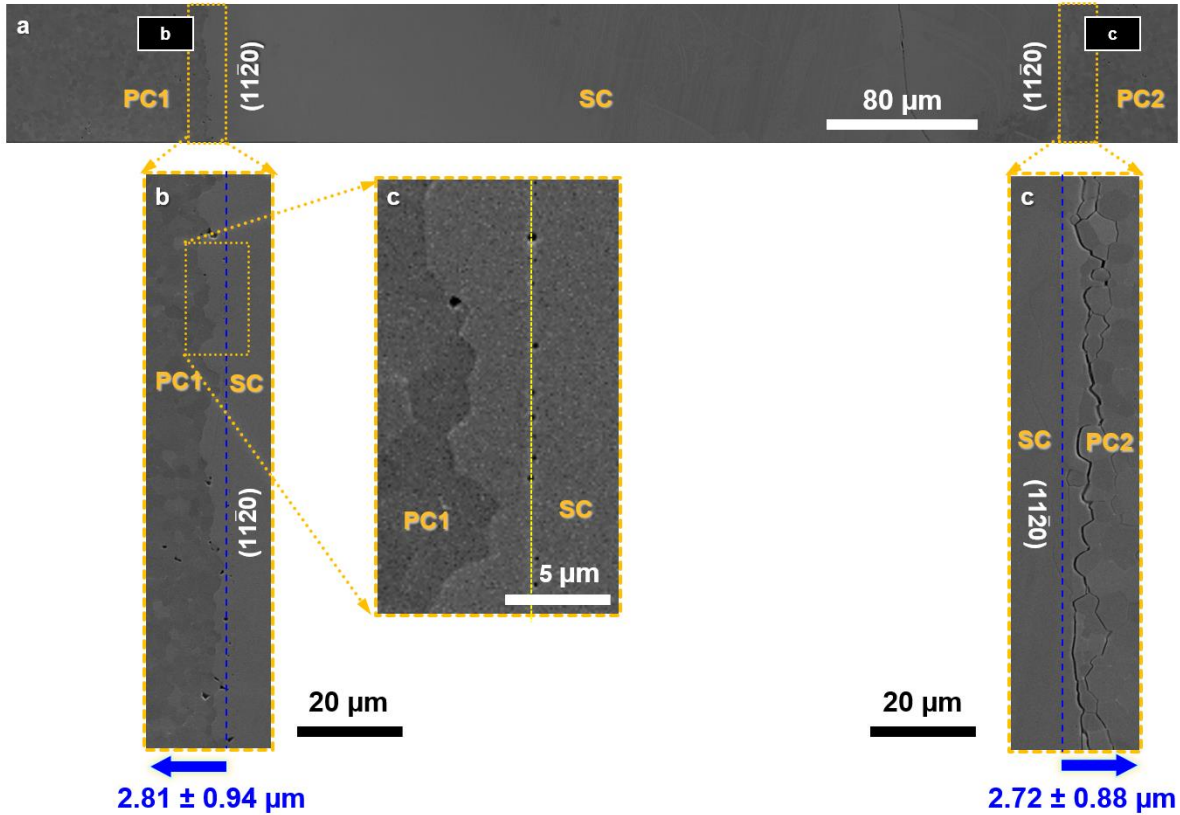

**Supplementary Fig. 6. Microstructure of an as-sintered Polycrystal 1/Single Crystal/Polycrystal 2 (PC1/SC/PC2) sandwich specimen. a**, Cross-sectional SEM micrograph of this $Bi_2O_3$-doped ZnO PC1/SC/PC2 sandwich specimen densified using SPS at 780 °C for 5 minutes under 50 MPa pressure, followed by de-carbonization annealing in air at 700 °C for 9 hours. **b**, **c**, High magnification SEM micrographs of the SC/PC interfaces at both sides. Aligned holes were used to identify the locations of the original surfaces of the single crystal, as shown in an enlarged image in Panel (c). The migration distances of the SC fronts on both sides are similar, which measured to be ~2.7-2.8 μm on average.

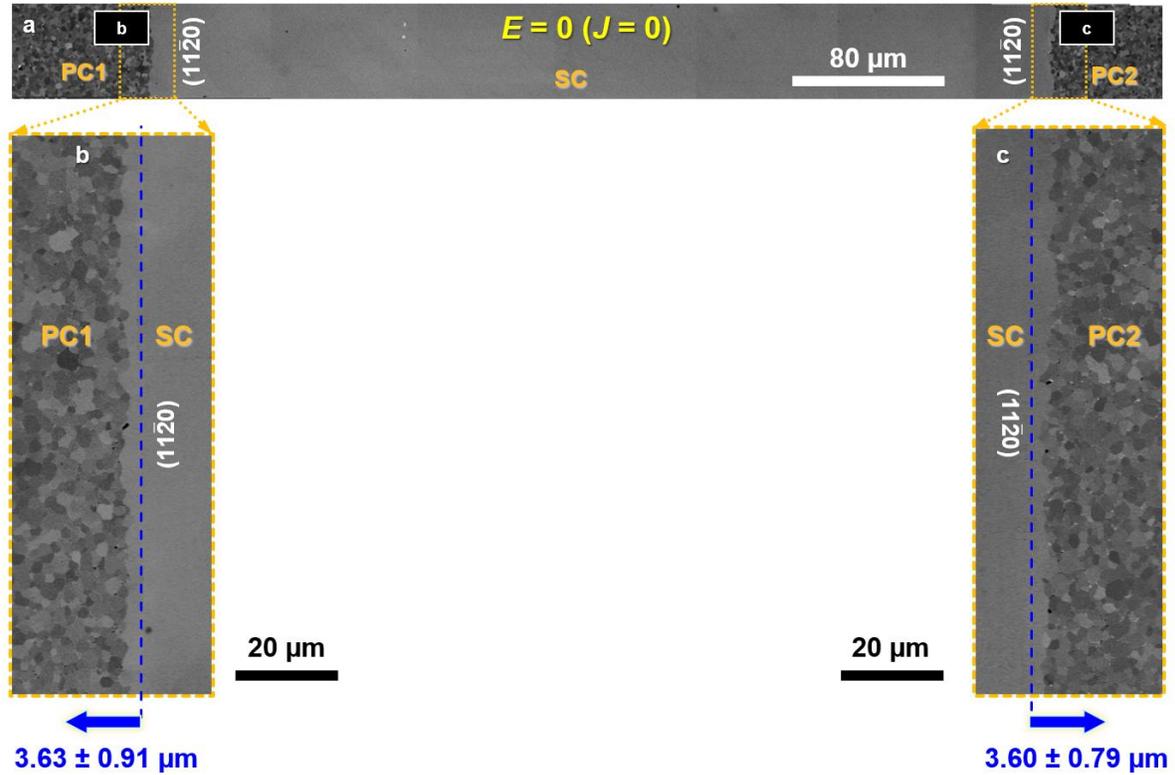

**Supplementary Fig. 7. Microstructure of a reference $Bi_2O_3$-doped ZnO PC1/SC/PC2 sandwich specimen annealed without an electric field isothermally at 880 °C for 4 hours and subsequently quenched. a**, Cross-sectional SEM micrograph of this PC1/SC/PC2 specimen. **b, c**, High-magnification SEM micrographs of the SC1/PC and SC/PC2 interfaces. Blue dashed lines indicate the original interfaces. The migration distances of the single crystal fronts on both sides are similar, which measured to be ~3.6 μm on average. Note that the migration distances here include the growths of the single crystal during the isothermal annealing as well as in prior fabrication steps (~2.7-2.8 μm on average during SPS and decarbonization annealing, as shown in Supplementary Fig. 6).

## Supplementary Note 3:

**Defects polarization and the profile of oxygen vacancies in the sandwich specimen**

The Fig. 2h schematically illustrates the Polycrystal 1/Single Crystal/Polycrystal 2 (PC1/SC/PC2) specimen, where with an applied current created the reduced PC1- and PC2- and oxidized PC1+ and PC2+ regions.

It is well known that ZnO single crystal is an electronic conductor (semiconductor) [56], but $Bi_2O_3$ is an oxygen ionic conductor [57]. Interestingly, the $Bi_2O_3$-enriched GBs (liquid-like IGFs) in $Bi_2O_3$-doped ZnO can be ion-conducting. To prove its ionic conductivity, our measurement showed a high ionic transfer number of ~0.9 in the polycrystalline $Bi_2O_3$-doped ZnO at 840 °C; see Supplementary Fig. 1c, which is also shown below for quick access.

Thus, the two PC regions in the PC1/SC/PC2 sandwich specimen are mostly oxygen ion conducting, separated by the electron-conducting but ion-blocking SC. Thus, an applied electric current in the PC/SC/PC should create two reduced regions PC1- and PC2- and two oxidized regions PC1+ and PC2+ with discontinuous oxygen chemical potentials at the PC1/SC and SC/PC2 interfaces, as schematically shown in Fig. 2h.

- On the one hand, the formation of the reduced regions with increased concentration of oxygen vacancies was proved directly by photoluminance spectroscopy (discussed in Supplementary Note 4 subsequently).
- On the other hand, the presence of the oxidized regions was indicated by the generation of pores (with the oxidation reaction: $O^{2-} \rightarrow \frac{1}{2}O_2 + e^-$) as evident, *e.g.*, the SC/PC1+ region in Fig. 2d (which are provided below for quick access) and the PC2+ region (as shown in the SEM image on the right side below).

Here, we can use a model to explain the defects polarization and the profile of oxygen vacancies shown in and Fig. 2h. Let us assume oxygen vacancies ($V_O^{\bullet\bullet}$ in the Kröger-Vink notation) as the main defects (similar results can be extended to other types of charged defects). On the one hand, the carriers with positive charge ($V_O^{\bullet\bullet}$) will move along the direction of the electric field and accumulate at the SC/PC2- interface (since the SC is ion blocking) as well as PC1- region near the cathode (blocking Pt electrode), thereby forming two locally reduced regions (Fig. 2h). On the other hand, PC1+ and PC2+ regions should remain oxidized (mostly stoichiometric, with minimum oxygen vacancies), where the oxidation reaction ($O^{2-} \rightarrow \frac{1}{2}O_2 + e^-$) occurs as evident by the formation of porosity (see, *e.g.*, Fig. 2d).

Subsequently, we can plot the schematic profiles of electric potential ($\phi$), chemical potential ($\mu_{O^{2-}}$), and electrochemical potential ($\eta_{O^{2-}}$) across the sandwich specimen in Fig. 2h, following a similar model from prior electrochemical studies of solid-oxide fuel cells (SOFCs) [38]. Different from the typical SOFC setups [38], our sandwich specimen was annealed in air (both sides) so that schematic profiles shown in Fig. 2h represent the middle part of the specimen (in the cross-sectional direction perpendicular to the applied electric field/current); consistently, there was more reduction in the middle section of the PC2- region that increased GB mobility, as shown in Supplementary Note 2.

Notably, the chemical potential $\mu_{O^{2-}}$ decreases nonlinearly along the direction of electric field in the polycrystal regions, with an abrupt jump (so called "oxygen potential transition" demonstrated by Chen and co-workers [38], a character of a mixed electron-ion conductor). In addition, the electric potential $\phi$ should also change abruptly in the polycrystals (as shown in the model proposed by Chen and co-workers [38]), but transit continuously at the PC/SC interfaces. Specifically, these profiles should be determined by applying the Wagner-type transport theory [58]:

$$\frac{d\phi}{dx} = -\frac{i_{tot}}{\sigma_{tot}} + \frac{t_{O^{2-}}}{2e}\frac{d\mu_{O^{2-}}}{dx} + \frac{t_{e^-}}{e}\frac{d\mu_{e^-}}{dx}, \qquad (1)$$

where $i_{tot}$ is the total current density (flux), $\sigma_{tot}$ is the total (ionic plus electronic) conductivity, $e$ is the charge of electron, and $t_{O^{2-}}$ and $t_{e^-}$ are the transport numbers of oxygen ions and electrons, respectively. The electrochemical potential is given by:

$$\eta_{O^{2-}} = \mu_{O^{2-}} - 2e\phi \qquad (2)$$

To solve equation (1), we may assume that (*i*) local chemical equilibria in the PC1+ and PC2+ regions with the reaction: $O^{2-} = \frac{1}{2}O_2 + 2e^-$, and (*ii*) $\Delta\mu_e$ equals to 0 with the blocking Pt electrodes. Then, the electrochemical potential of $O^{2-}$ can be obtained:

$$\frac{\partial \eta_{O^{2-}}}{\partial x} = \frac{2ei_{tot.}}{\sigma_{tot.}} + (1 - t_O)\frac{\partial \mu_{O^{2-}}}{\partial x} \qquad (3)$$

Based on equations (1-3) and the above assessments, the electric, chemical, and electrochemical potential profiles can be schematically plotted in Fig. 2h. Unfortunately, we cannot quantify these profiles because most of the parameters used in equations (1-3) are unknown. Yet, it can be concluded that an applied current through the PC1/SC/PC2 sandwich would create the reduced PC1- and PC2- and oxidized PC1+ and PC2+ regions because the SC is ion-blocking and the PC1 and PC2 regions are ion-conducting (with a measured ionic transfer number of ~0.9, as shown in Supplementary Fig. 1c).

# Supplementary Note 4:

## Analysis of photoluminescence spectra and oxygen vacancy concentrations

Photoluminescence spectroscopy was used to analyze oxygen defects in the ZnO sandwich specimen induced by an applied electric current (Fig. 3).

A typical photoluminescence spectrum of ZnO has two main peaks associated with the oxygen defects in the (*i*) UV emission (at ~3.25 eV) due to exciton-bound states and (*ii*) visible light emission (at ~2.15 eV) due to deep-level states [59,60]. Since oxygen vacancies act as deep-donor defects [61], we collected photoluminescence signals in the visible light region from 400 to 700 nm (see, *e.g.*, Supplementary Fig. 8b, c).

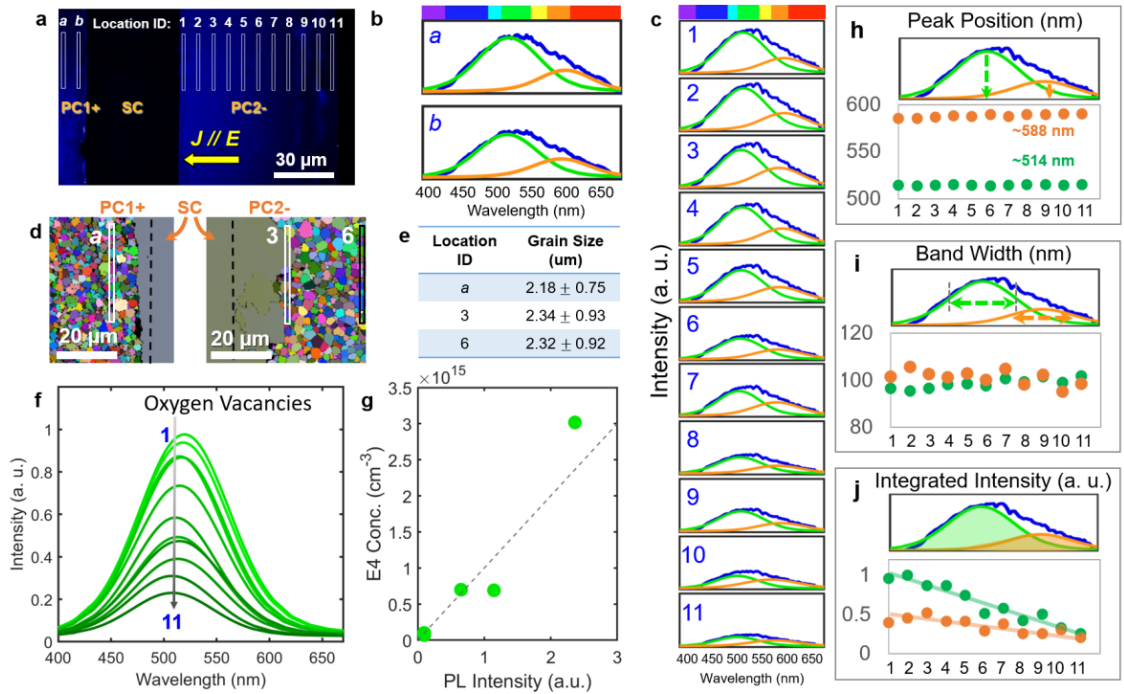

**Supplementary Fig. 8. Band decomposition and analysis of the photoluminescence spectra of the $Bi_2O_3$-doped ZnO PC1/SC/PC2 sandwich specimen annealed with a constant applied electric current, showing the enrichment of oxygen defects in the reduced PC2- region. a**, Photoluminescence intensity maps at the 526 nm wavelength of the cross-sectional PC1/SC/PC2 specimen. **b**, Band decomposition of photoluminescence spectra in the PC1+ region collected at Locations *a* and *b* labeled in Panel (a). **c**, Band decomposition of photoluminescence spectra in the PC2- region collected at Locations 1-11 labeled in Panel (a). The photoluminescence bands can be decomposed to two overlapping emission bands. The "green" band in Panel (b) and (c) is related to the emission from oxygen vacancies[61]. **d**, The EBSD maps of the PC1+ and PC2- regions. **e**, Measured grain sizes of polycrystals near the interfaces with the single crystal (at Location *a* in the PC1+ region and Location 3 in the PC2- region), as well as at Location 6 in the PC2 away from the SC/PC2- interface, showing no substantial differences in the grain sizes; thus, the observed variations in the photoluminescence intensities are not due to the grain size effects. **f**, The evolution of the green-emission bands (representing oxygen vacancies) from Location 1 to 11 (away from the SC/PC2- interface). **g**, The concentration of the so-called "E4 defects" (corresponding to the 4th energy level below the conduction band; attributed to oxygen vacancies based on DFT calculations[61-63]) *vs.* the green-emission band intensity at 2.45 eV (~506 nm) measured by deep level transient spectroscopy (DLTS)[61]. The decreasing of green-emission band intensity from Location 1 to 11, as shown in Panel (f), indicates decreasing oxygen vacancy concentrations. **h-j**, The peak position, band width, and integrated intensity of the separated "green" and "orange" bands at Locations 1-11.

Subsequently, we decomposed photoluminescence spectra into "green" (centered at ~510 nm) and "orange" (centered at ~600 nm) bands to further analyze the defects, particularly the distribution of oxygen vacancies. It is worth noting that the two decomposed bands are named as "green" and "orange" because their center wavelengths are in the green and orange spectra, respectively, albeit that both peaks are broad and overlap significantly. It is well established from prior studies that the presence of oxygen vacancies dominantly contribute to the green-band emission around 510 nm [59,61,64,65]. Most prior studies considered orange-band emission to be related to zinc interstitials [65,66], albeit a debate [61].

In this study, we focused on the decomposed "green" bands to probe oxygen vacancy distribution. Hoffman *et al.* used deep level transient spectroscopy (DLTS) to find a so-called "E4 defect level" that is ~0.53 eV below conduction band (CB), similar to the defect level of the oxygen vacancy (*i.e.*, ~0.6 eV below CB based on first-principles calculations); thus, they attributed this "E4 defect" (*a.k.a.* the 4$^{th}$ defect energy level below the CB) to oxygen vacancy [61]. They also found that the concentration of "E4 defects" (oxygen vacancies) is proportional to the photoluminescence intensity (Supplementary Fig. 8) so that the green-band intensity should have a positive correlation with the oxygen vacancy concentration [61]. Thus, we used the intensity of this decomposed green band to probe the distribution of oxygen vacancies in this study.

To quantify the intensity of the green-band emission, we used the CasaXPS software to fit decomposed bands by using the asymmetric Lorentzian lineshape method with Gaussian distribution. The lineshape parameters were set to the default values: $\alpha = 1.53$ and $\beta = 243$. The band positions were constrained in the range of 415 to 670 nm.

The decomposed green and orange bands are plotted for Locations *a* and *b* in Supplementary Fig. 8a, b and for Locations 1-11 in Supplementary Fig. 8c. The fitted peak positions and band widths for the two bands remain mostly constant (Supplementary Fig. 8h, i), while the integrated intensity of the green band, which represents in the oxygen vacancy concentration as discussed above, decreases from Location 1 to Location 11 (*i.e.*, moving away from the SC/PC2- interface), as shown in Supplementary Fig. 8j, f. The intensity of the green band is also lower at the PC1+ region (Location *a*), as shown in Fig. 3e. This suggests a higher concentration of oxygen vacancies near the SC/PC2- interface.

The photoluminescence emission peak was not observed in the single crystal (SC) region, thereby suggesting the defects are mostly present at GBs. It is in fact well known that photoluminescence peaks associated with oxygen defects are mostly from GBs in ZnO, so that the grain size may also have an effect on photoluminescence intensity [59,64]. Here, we need to exclude this effect to draw a rigorous conclusion. Our measurements based on EBSD maps show the reduced PC2- region actually has slightly larger mean grain size (2.34 ± 0.93 μm) than the oxidized PC1+ (2.18 ± 0.75 μm) and the grain size does not change significantly (virtually a constant within ~50 μm) in the PC2- region (*e.g.*, 2.34 ± 0.93 μm at Location 3 and 2.32 ± 0.92 μm at Location 6), as shown in Supplementary Fig. 8e. Thus, the decreasing intensity of the green band away from the SC/PC2- interface in the PC2- region (Supplementary Fig. 8f), as well as the higher intensity in the (reduced) PC2- region than that in the (oxidized) PC1+ region (Fig. 3c), is not due to the grain size effect. Instead, the PL analysis directly showed a higher concentration of oxygen vacancies in the reduced PC2- region, particularly near the SC/PC2- interface.

Notably, the intensity of the green-band emission in the PC2- region is higher than that in the PC1+ region (as shown in Fig. 3e). This directly revealed the presence of more oxygen vacancies in the reduced SC/PC2- interface (*i.e.*, the electrochemically reduced region where GB disorder-to-order transition led to abruptly increased GB mobilities) than the oxidized PC1+/SC interface (with the disordered and slow-moving GBs).

# Supplementary Note 5:
## Amorphous-like *vs.* ordered GB structures and EDS confirmation of Bi segregation

<u>Amorphous-like *vs.* Ordered GB Structures</u>: Selected representative AC STEM high-angle annular dark-field (HAADF) and bright field (BF) images shown in Fig. 1 suggest that electrochemical reduction induced a GB disorder-to-order transition, which subsequently led to abruptly increased GB mobilities.

Key observations of slow-moving disordered GBs vs. fast-moving ordered GBs in four cases are summarized as follows:

- Amorphous-like GBs, commonly known as intergranular "glassy" films (IGFs, albeit the existence of partial orders [68,69]), have been observed in the reference specimen without an applied electric field/current (Supplementary Fig. 9 and Fig. 1a, b).
    - Such nanometer-thick IGFs, which are ubiquitously observed in sintered ceramics [70], can alternatively be understood to be liquid-like interfacial films that adopt an "equilibrium" thickness on the order of 1 nm in response to a balance of attractive and repulsive interfacial interactions (originally proposed by Clarke [71]) or disordered multilayer adsorbates (originally proposed by Cannon *et al.* [72]).
    - Specifically, such nanoscale liquid-like IGFs have been observed to form at all general GBs (examined to date) in $Bi_2O_3$-sataured ZnO at thermodynamic equilibria both above and below the bulk eutectic temperature (740 °C), as well as some un-saturated specimens below the bulk solid solubility limits [67,73,74].
- Amorphous-like (disordered) GBs or nanoscale IGFs of similar character have also been observed at the oxidized, slow-moving PC1+/SC interface (Supplementary Fig. 10 and Fig. 1e, f) in the specimen with an applied electric current, which exhibit a similar mobility as the PC/SC interfaces in the reference specimen without an applied electric field.
- In contrast, ordered GB complexions (ordered Bi adsorbates) have been observed at the reduced, fast-moving SC/PC2- interface (Supplementary Fig. 11 and Fig. 1g, h) in the specimen with an applied electric current.
- In addition, ordered Bi adsorbates have also been observed at the GBs of the fast-moving abnormal grains in the reduced PC1- region (Supplementary Fig. 12 and Fig. 1c-d) in the specimen with an applied electric current.

In summary, the slow-moving GBs in the reference specimen without an applied electric field/current and in the oxidized region of the specimen with an applied electric current are all disordered (forming nanoscale amorphous-like or liquid-like IGFs), while the fast-moving GBs in electrochemically reduced regions are all ordered (albeit different ordered structures, which are presumably due to the different crystallographic GB characters of at different general GBs randomly selected from the specimen). Representative AC STEM images are shown in Fig. 1; additional images and cases are documented in Supplementary Figs. 9-12 show the generality of the observations slow-moving disordered GBs *vs.* fast-moving ordered GBs.

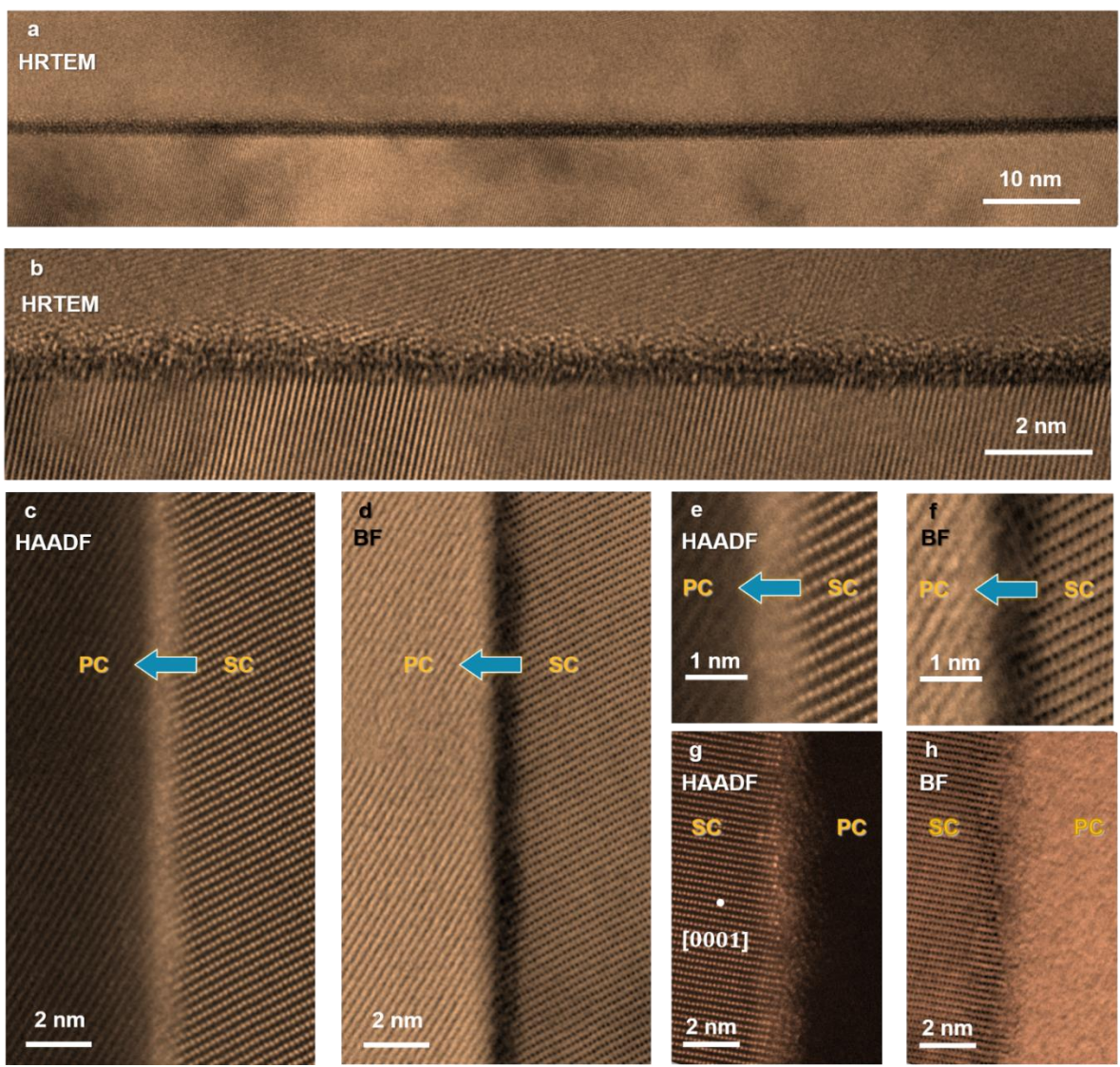

**Supplementary Fig. 9. High-resolution transmission electron microscopy (HRTEM) and AC STEM images of the (slow-moving) disordered GBs in Bi$_2$O$_3$-doped ZnO specimens annealed at 880°C with no electric field. a, b**, HRTEM images and **c-h**, AC STEM HAADF and BF images of the characteristic "amorphous-like" intergranular films (IGFs) (*a.k.a.* "disordered" GBs) in Bi$_2$O$_3$-doped ZnO specimens quenched from 880 °C after annealing for 4 hours with no electric field. Panels (e) and (f) are an enlarged view of the GB shown in Fig. 1a, b and all others are additional images to show the ubiquitous formation of such disordered GBs in Bi$_2$O$_3$-doped ZnO. Such nanoscale amorphous-like IGFs have been observed to form at all general GBs in Bi$_2$O$_3$-saturated ZnO at thermodynamic equilibria both above and below the bulk eutectic temperature (740 °C) in prior studies[67,73,74].

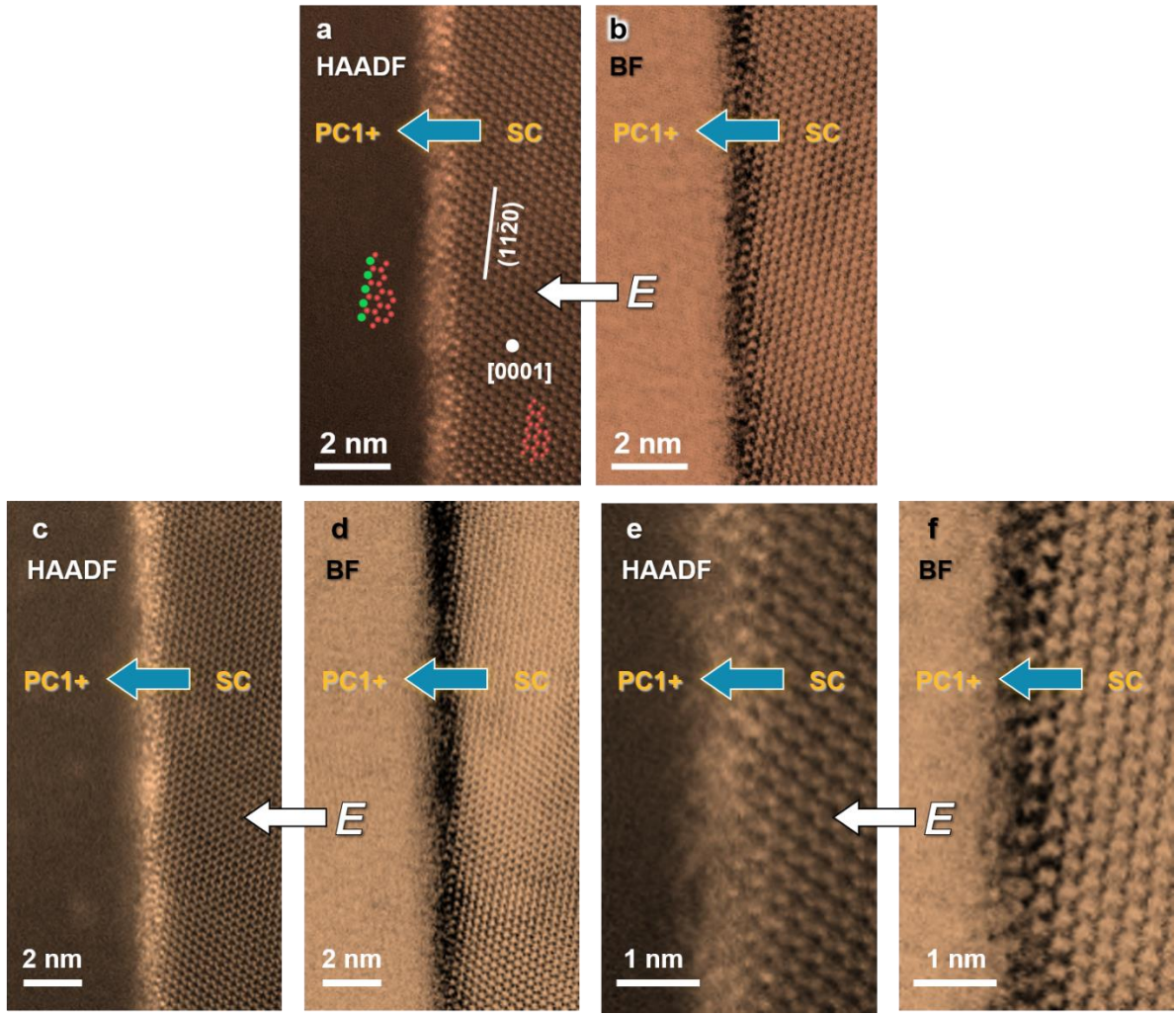

**Supplementary Fig. 10. HAADF and BF STEM images of the (slow-moving) disordered GBs in the oxidized PC1+ region.** Panels (a) and (b) are an enlarged view of the GB shown in Fig. 1e, f and all others are additional AC-STEM images showing the ubiquitous observations of disordered GBs of similar character in the oxidized PC1+ region of the $Bi_2O_3$-doped ZnO sandwich specimen annealed with constant $J = 6.4$ mA/mm$^2$.

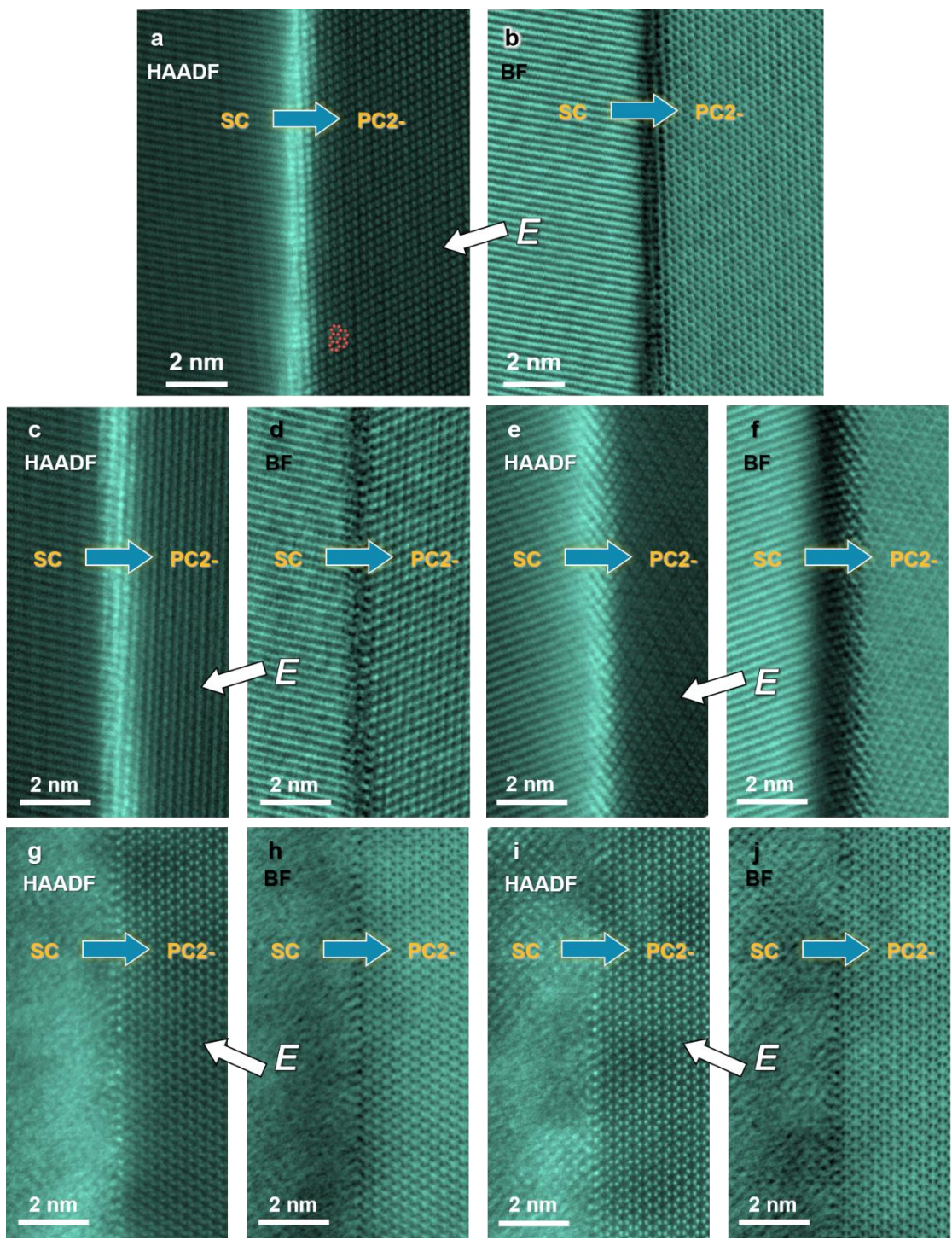

**Supplementary Fig. 11. HAADF and BF STEM images of the (fast-moving) ordered GBs in the reduced PC2- region.** STEM images of representative (**a-f**) bilayer- or multilayer-like and (**g-j**) monolayer-like ordered GB complexions. This specimen was annealed with constant 6.4 mA/mm$^2$. Panels (a) and (b) are a larger view of the GB shown in Fig. 1g, h and all others are more AC-STEM images showing the ubiquitous observations of ordered GBs in the reduced PC2- region of the ZnO-Bi$_2$O$_3$ sandwich specimen. The fast-moving ordered GB did not maintain the initial orientation; the images are rotated so that the GBs are vertical, where the field directions are indicated by the arrows.

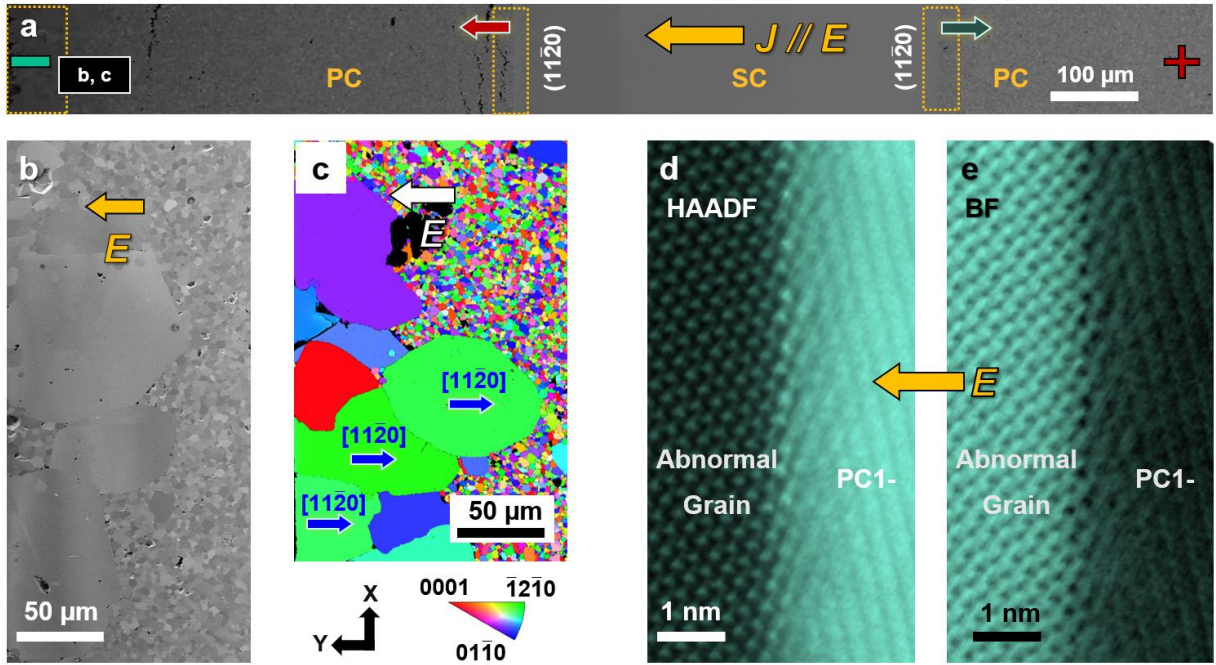

**Supplementary Fig. 12. Microstructure and crystallography of the abnormal grains near the cathode in the reduced PC1- region in the sandwich specimen and EBSD map of an additional polycrystalline specimen, both annealed with applied electric currents. a**, **b**, Cross-sectional SEM of the abnormal grains near the cathode of this specimen. **c,** Y-direction inverse pole figure (Y-IPF) in EBSD micrographs of the abnormal grains. **d**, **e**, AC HAADF and BF STEM images of a fast-moving, ordered GB between an abnormal grain and an abutting smaller (normal) grain. This specimen was annealed with constant $J$ = 6.4 mA/mm$^2$.

EDS Confirmation of Bi Segregation: The bright contrasts at GBs in the HAADF images are due to heavy Bi adsorbates (because of the $Z$ contrast). Bi segregation has also been directly verified by energy dispersive X-ray spectroscopy (EDS) analysis (Supplementary Fig. 13). Note that this GB was intentionally tilted slightly to reduce beam damages (i.e., the GB is not exactly edge-on, so that the Bi segregation region looks wider in the Bi elemental map).

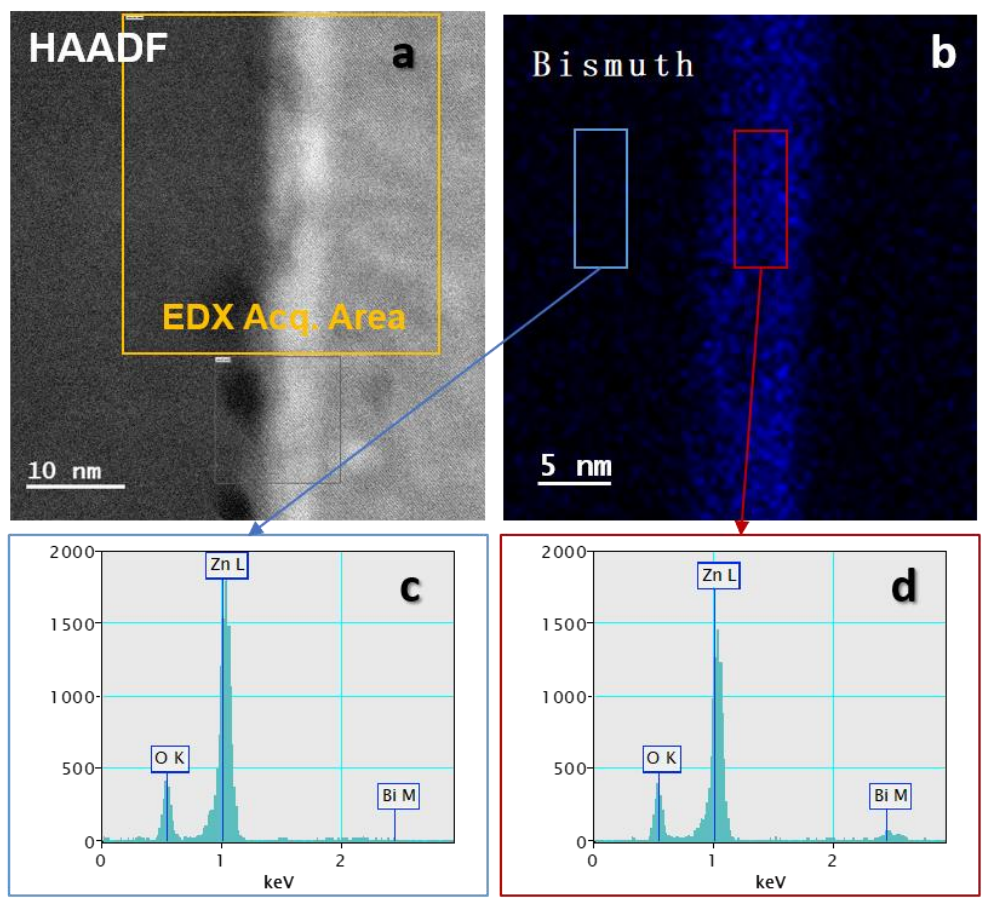

**Supplementary Fig. 13. Energy dispersive X-ray spectroscopy (EDS) confirming the GB segregation of Bi. a** STEM HAADF image, **b**, EDS Bi elemental map, and EDS spectra of two selected areas **c,** inside the ZnO grain and **d,** at the Bi-enriched GB, respectively. This GB was in the reduced PC1- region in the $Bi_2O_3$-doped ZnO specimen quenched after annealing for 4 hours with the constant current density of 6.4 mA/mm$^2$. Noting that this GB was intentionally tilted slightly to reduce beam damages (*i.e.*, the GB is not exactly edge-on, so that the Bi segregation region looks wider in the Bi elemental map). It is known that $Bi_2O_3$ adsorbates can be damaged (and presumably reduced) rapidly by the irradiation of an intense electron beam in high vacuum in a TEM (as shown for analogous surface phase in $Bi_2O_3$-doped ZnO in a prior study [83]).

## Supplementary Note 6:
### DFT optimized GB structures and comparisons with experiments

The procedure for constructing GB models and the details of the density functional theory (DFT) calculations can be found in Methods. A few key points are summarized as follows. A low-symmetry GB model (containing 240 atoms) was selected to represent general GBs, where one terminal GB plane was set to be $(11\bar{2}0)$ to best mimic the SC/PC interface in our experiment within the size limit of the DFT calculations. To dope stoichiometric GBs, we added one oxygen atom for every two Bi dopants (normally forming two $Bi^{3+}$ cations to replace two $Zn^{2+}$ cations) to keep the charge neutrality and fully relaxed the GB structures. In most calculations, the amount of GB excess of Bi was selected to match the experimentally measured average value from prior studies [67,74]. In each case, the GB free volume was allowed to be relaxed to achieve an equilibrium.

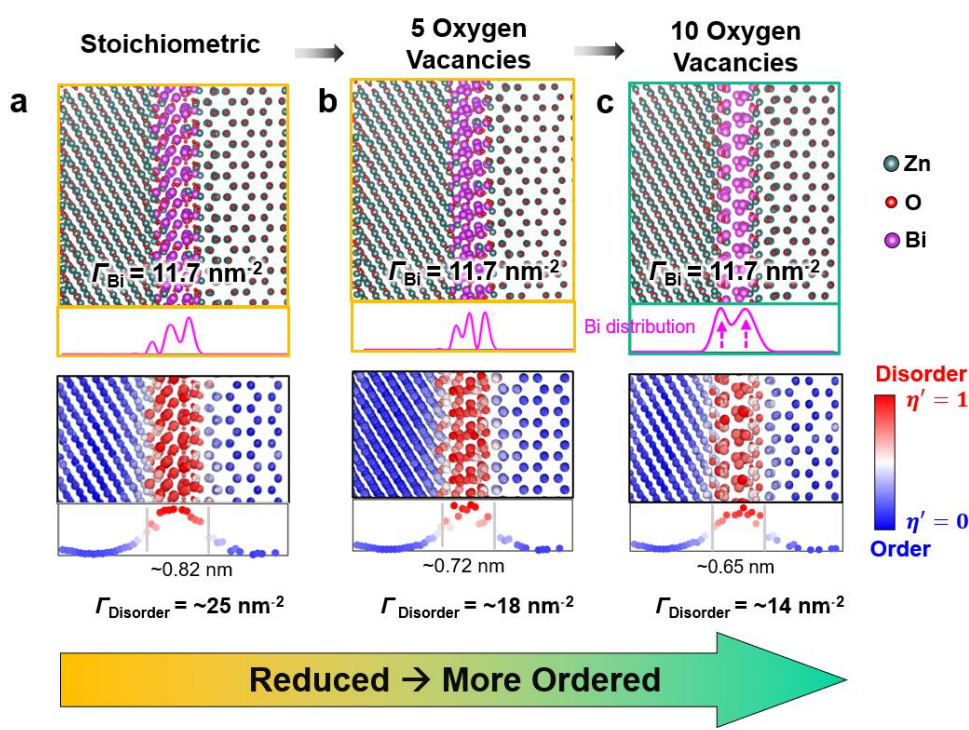

**Supplementary Fig. 14. Variation of GB structures with increasing levels of reduction predicted by DFT. a**, DFT-relaxed stoichiometric ZnO-$Bi_2O_3$ GB structure with a $\Gamma_{Bi}$ of 11.7 Bi atoms $nm^{-2}$ (equivalent to about one monolayer of Bi coverage, which is about the averaged value from prior STEM EDS measurements [55]). Note that one oxygen atom is added into the GB region after doping every two Bi atoms (to replace two Zn atoms) to keep the stoichiometry. **a**, The DFT-relaxed stoichiometric GB (mimic the oxidized condition in experiments). The DFT-relaxed reduced GB structures by (**b**) removing ~5.8 oxygen atoms per $nm^2$ (about half monolayer) and (**c**) ~11.7 oxygen atoms per $nm^2$ (about one monolayer), respectively. The computed Bi distribution and disorder profiles indicate that oxygen reduction leads to the formation of more ordered GB structures.

To simulate reduced GBs, we removed a controlled number of oxygen atoms and fully relaxed the interfacial structure again. A series of DFT optimized interfacial structures of stoichiometric and reduced GBs are shown in Supplementary Fig. 14, which reveal that reduction can make the GB more ordered. To better quantify the GB disorder, we calculated a dimensionless disorder parameter ($\eta' = 1$ for an atom in a liquid and $\eta' = 0$ for an atom in a perfect crystal; $\eta' = 1 - \eta$, where $\eta$ is an order parameter; see Methods for the definition and quantification method).

Noting that the DFT-relaxed stoichiometric GB represent the GB observed in the oxidized conditions in experiments (since we expect that no excess oxygen at GBs in experiments).

Then, we computed the GB excess disorder ($\Gamma_{\text{Disorder}}$) by integrating the $\eta'(x)$ profile for a GB located at $x = 0$ based on the methods used in prior studies [75-78].

As shown in Supplementary Fig. 14, the GB excess disorder decreased from $\Gamma_{\text{Disorder}} = \sim 25$ nm$^{-2}$ for the stoichiometric amorphous-like GB to $\Gamma_{\text{Disorder}} = \sim 18$ nm$^{-2}$ after removing about half monolayer of oxygen atoms (*i.e.*, ~5.8 oxygen atoms per nm$^2$) and $\Gamma_{\text{Disorder}} = \sim 14$ nm$^{-2}$ after removing about one monolayer of oxygen atoms (*i.e.*, ~11.7 oxygen atoms per nm$^2$), respectively.

Specifically, the reduced GB shown in Supplementary Fig. 14c (after removing about one monolayer of oxygen atoms) exhibit a characteristic interfacial structure of bilayer-like Bi adsorption resembling that observed at the SC/PC2- interface by AC STEM in Fig. 5b in the main article. Thus, we show this ordered (bilayer-like) reduced GB along with the amorphous-like GB in Fig. 5a and critically compare them with STEM HAADF images.

Specifically, Fig. 5a-f in the main article compares experimentally observed and DFT simulated stoichiometric (disordered) *vs.* reduced (ordered) GB structures:
- Expanded views of STEM HAADF images for a stoichiometric and disordered GB *vs.* a reduced and ordered (bilayer-like) GB are shown in Fig. 5a *vs.* 4b, where the averaged intensity and line-by-line FFT patterns are also plotted to illustrate the layering and periodic orders. Here, DFT-relaxed stoichiometric GB represent the GB observed in the oxidized conditions in experiments (as discussed above).
- The STEM images are compared with DFT-optimized structures of the stoichiometric *vs.* reduced GBs shown in Fig. 5c *vs.* 4d, where the Bi concentration profiles projected along the *x* direction are plotted beneath.
- The calculated disorder parameters for all atoms for DFT-optimized stoichiometric *vs.* reduced GB structures are shown in Fig. 5e *vs.* 4f. The disorder parameter profiles $\eta'(x)$ projected along the *x* direction are plotted above.

Overall, DFT calculations agree well with the experimental observations:
- On the one hand, the stoichiometric GB is disordered, as shown in the STEM image and FFT analysis in Fig. 5a, with a computed $\Gamma_{\text{Disorder}}$ of ~25 nm$^{-2}$. The width of the disordered layer (IGF) was measured to be ~0.9 nm from the STEM HAADF image shown in Fig. 5a (and ~0.7-0.9 nm for different IGFs observed in this study) and ~0.8 nm from the DFT simulated disorder profile (Fig. 5e), which agree with each other.
- On the other hand, the reduced GB is more ordered with a bilayer-like structure shown in

both the STEM image (Fig. 5b) and the Bi adsorption profile obtained by the DFT optimization (Fig. 5c), with a reduced computed $\Gamma_{\text{Disorder}}$ of ~14 nm$^{-2}$. Both the STEM HAADF image and DFT calculations show that the interlayer distance (within the bilayer-like interfacial structure) is ~0.3 nm.

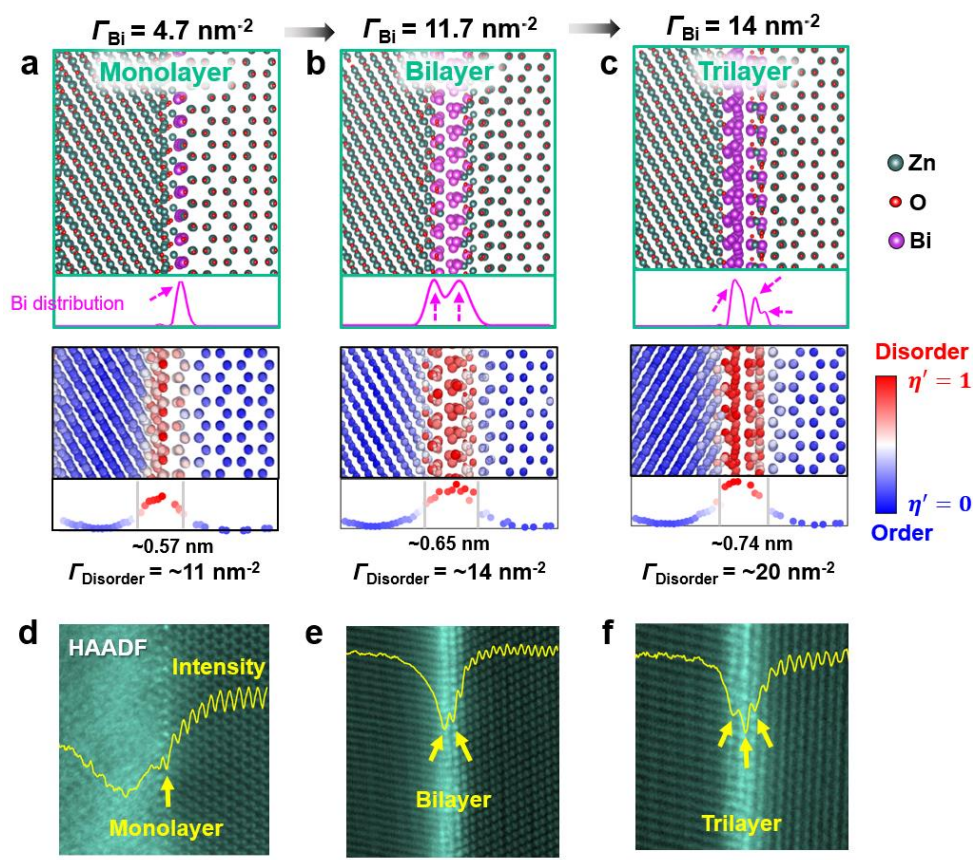

**Supplementary Fig. 15. A series of ordered complexions (Bi adsorption structures) formed at electrochemically reduced GBs. a-c**, DFT-optimized reduced GB structures with increasing Bi adsorption ($\Gamma_{\text{Bi}}$), showing the formation of monolayer-, bilayer-, and trilayer-like complexions. **d-f**, STEM HAADF images of ordered GBs observed in the reduced regions of our specimen that resemble these layered adsorption structures predicted from DFT (albeit different GB characters). It is noted that experimental observations were made on GBs randomly-selected from the reduced region, which cannot be modeled accurately due the lacking of full details of the GB crystallographic characters and the limitations of the maximum size for effective DFT calculations.

In addition to GB bilayer-like structure, DFT calculations show that reduced ZnO GBs can also exhibit a range of ordered segregation structures, from monolayer to trilayer with increasing levels of GB disorder and interfacial width Supplementary Fig. 15). Interestingly, similar interfacial structures have also been observed in randomly selected general GBs in the electrochemically reduced regions in STEM HAADF images (Supplementary Fig. 15d-f) in this study. More critical assessments are not warranted here because we could not determine the exact crystallographic characters of randomly-selected general GBs from experiments in most cases to build cells to allow DFT calculations (which are also limited for most general GBs because realistic models often contain too many atoms). Nevertheless, these experimental discoveries, along with DFT modeling, demonstrate the generality of forming ordered GB structures in the reduced regions of $Bi_2O_3$-doped ZnO.

In summary, DFT calculations have confirmed that the reduction can induce a GB disorder-to-order transition in $Bi_2O_3$-doped ZnO, and the predictions are consistent with experiments.

# Supplementary Note 7:

## DFT calculations of GB energies and a predicted GB disorder-order transition

To further verify and model the reduction-induced GB disorder-order transition, we calculated the GB energy as a function of chemical potentials for both stoichiometric and reduced GBs. The GB energy ($\gamma_{GB}$) can be determined from DFT calculations using the following equation:

$$\gamma_{GB} = \frac{E_{Total}^{GB\ Bi@ZnO} - E_{Grain1}^{ZnO} - E_{Grain2}^{ZnO} - n_O \mu_O + n_{Bi/Zn}\mu_{Zn} - n_{Bi/Zn}\mu_{Bi}}{A}, \quad (4)$$

where $E_{Total}^{GB\ Bi@ZnO}$ is the total energy of the Bi-doped ZnO DFT calculation cell with the GB (after the full relaxation), $E_{Grain1(2)}^{ZnO}$ is the reference energy of Grain 1 (or 2), $n_{Bi/Zn}$ is the number of Zn atoms substituted by Bi atoms, $n_O$ is the number of excess oxygen atoms (*i.e.*, number of the oxygen atoms added at the GB to compensate the charge of aliovalent Bi doping minus the number of the oxygen atoms removed for creating reduction), $\mu_i$ ($i$ = O, Zn, or Bi) is the chemical potential of O, Zn, or Bi, and $A$ is the cross-sectional area of the GB. The oxygen chemical potential difference $\Delta\mu_O$ can be defined as:

$$\Delta\mu_O = \mu_O - \frac{1}{2}E_{O_2}, \quad (5)$$

where $E_{O_2}$ is the energy of the oxygen gas molecule with a correction [82]. The boundaries of $\Delta\mu_O$ are given by:

$$\Delta\mu_O < 0, \quad (6)$$

and

$$\Delta\mu_O + \Delta\mu_{Zn} = \Delta H_f(ZnO), \quad (7)$$

where $\Delta H_f(ZnO)$ is the formation enthalpy of ZnO. Similarly, we define the Bi chemical potential difference as:

$$\Delta\mu_{Bi} = \mu_{Bi} - E_{Bi}, \quad (8)$$

where $E_{Bi}$ is the DFT energy of the Bi element (in its stable solid form). The upper and lower boundaries of $\Delta\mu_{Bi}$ can be specified by a similar method.

Using equations (5-8), we calculated GB energies for both stoichiometric and reduced GBs. In the Supplementary Fig. 16, we plotted DFT-calculated GB energies $\gamma_{GB}$ of stoichiometric/disordered and reduced/ordered GBs as a function of chemical potential difference of oxygen ($\Delta\mu_O$) and bismuth ($\Delta\mu_{Bi}$), where the intersection line between the $\gamma_{GB}^{Reduced}$ and $\gamma_{GB}^{Stoichiometric}$ planes define a GB transition.

Subsequently, we define the GB energy difference, $\Delta\gamma_{GB}$, as:

$$\Delta\gamma_{GB} = \gamma_{GB}^{Reduced} - \gamma_{GB}^{Stoichiometric}, \quad (9)$$

where $\gamma_{GB}^{\text{Stoichiometric}}$ and $\gamma_{GB}^{\text{Reduced}}$, respectively, are the GB energies for stoichiometric (disordered) and reduced (ordered) GBs, respectively, calculated from DFT. Here, $\Delta\gamma_{GB} = 0$ defines a transition.

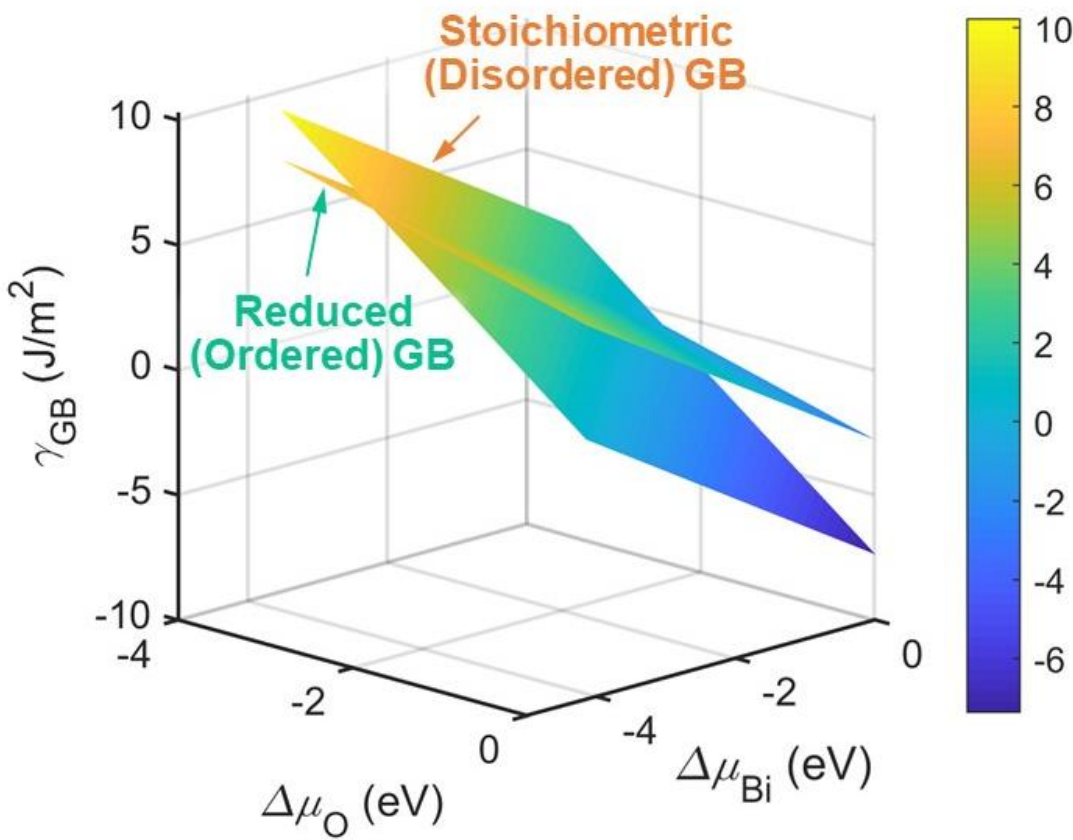

**Supplementary Fig. 16. DFT calculated GB energies to predict the transition between the stoichiometric and reduced GBs.** DFT calculated GB energies $\gamma_{GB}$ of stoichiometric (disordered) and reduced (ordered) GBs as a function of chemical potential difference of oxygen ($\Delta\mu_O$) and bismuth ($\Delta\mu_{Bi}$). Note that this calculation adopted same GB excess of Bi adsorption for both GBs, so that the difference in the GB energies $\Delta\gamma_{GB}$ (= $\gamma_{GB}^{\text{Reduced}} - \gamma_{GB}^{\text{Stoichiometric}}$) is independent of $\Delta\mu_{Bi}$. Consequently, a simplified plot of $\Delta\gamma_{GB}$ as a function of $\Delta\mu_O$ can be plotted in Fig. 6a.

Since both stoichiometric and reduced GBs have the same amount of GB excess of Bi, the difference in the GB energies $\Delta\gamma_{GB}$ is in fact independent of $\Delta\mu_{Bi}$. Thus, we can simplify Supplementary Fig. 16. Consequently, Fig. 6a shows a simplified plot of $\Delta\gamma_{GB}$ as a function of $\Delta\mu_O$ that represents this GB transition from a stoichiometric (disordered) GB to a reduced (ordered) GB.

# Supplementary Note 8:
## A generalizable thermodynamic model supported by DFT calculations

Here, we further discuss the physical origin of the reduction-induced GB disorder-to-order transition (*a.k.a.* why reduction can induce GB ordering) via a generalizable phenomenological thermodynamic model supported by DFT calculations.

Tang, Carter, and Cannon proposed a diffuse-interface model for a GB disorder-order transition, where the excess grand potential for a two-component GB is given by [79]:

$$\sigma^x = \int_{-\infty}^{+\infty} \left[ \Delta f(\eta, c) + \frac{\kappa_\eta^2}{2} \cdot \left(\frac{d\eta}{dx}\right)^2 + \frac{\kappa_c^2}{2} \cdot \left(\frac{dc}{dx}\right)^2 + s \cdot g(\eta) \cdot \left|\frac{d\theta}{dx}\right| \right] dx. \quad (10)$$

Here, the profiles of concentration $c(x)$, crystallinity $\eta(x)$ ($= 1 - \eta'(x)$, where $\eta'$ is the disorder parameter), and crystallographic orientation $\theta(x)$, are functions of $x$, the spatial parameter perpendicular to a GB located at $x = 0$; $\Delta f(c, \eta)$ is the homogenous (bulk) free energy density referred to the equilibrium bulk phases. Gradient energy coefficients, $\kappa_\eta$, $\kappa_c$, and $s$, are model parameters (materials constants). In equation (10), $g(\eta)$ is a function characterizing the coupling between $|d\theta/dx|$ and $\eta$, and we can assume that $g(\eta) = \eta^2$ (consistent with the Read-Shockley model) for simplicity [80]; high order terms can be included but this simple form can capture the basic underlying physics of the order-disorder transitions [79,81] so that it is adopted here. The equilibrium $c(x)$, $\eta(x)$, and $\theta(x)$ profiles should minimize the excess grand potential (*a.k.a.* interfacial energy) in equation (10).

Moreover, it was proven that $\theta(x)$ should take a step function [79,80]. Subsequently, minimization of the excess grand potential (interfacial energy) in equation (10) leads to [79]:

$$\sigma^x = s \cdot \Delta\theta \cdot \eta_{GB}^2 + \Delta F(\eta_{GB}), \quad (11)$$

where $\eta_{GB}$ is the order parameter at $x = 0$, $\Delta\theta$ is the GB misorientation, and

$$\Delta F(\eta_{GB}) \equiv \min \left\{ 2 \cdot \int_{\eta_{GB}}^{1} \sqrt{2\Delta f(\eta, c) \cdot \left[\kappa_\eta^2 + \kappa_c^2 \cdot \left(\frac{dc}{d\eta}\right)^2\right]} d\eta \right\}. \quad (12)$$

In equation (11), the first term $s \cdot \Delta\theta \cdot \eta_{GB}^2$ represents an energetic penalty to have a GB misorientation $\Delta\theta$, where GB disordering (or a smaller $\eta_{GB}$) lowers this energetic penalty. The second term $\Delta F(\eta_{GB})$ represents the total increased free energy due to GB disordering and associated compositional variation (*i.e.*, GB segregation) and gradient energy penalties (from the second and third terms in the integral in equation (10)) in a diffuse interface. From the definition in equation (12), we know:

$$\Delta F\left(\eta_{\text{GB}}=1\right)=0 \tag{13}$$

and

$$\Delta F\left(\eta_{\text{GB}}<1\right)>0. \tag{14}$$

In equation (12), the value of the integral also depends on the function $c(\eta)$ and $c_{\text{GB}}$, the composition at the GB. In a prior work [79], Tang, Carter, and Cannon proved (via a slightly different approach) that the following relations must be held for an equilibrium GB in a binary alloy to specify $\eta_{\text{GB}}^{\text{Equilibrium}}$ and $c_{\text{GB}}^{\text{Equilibrium}}$ at a thermodynamic equilibrium:

$$s \cdot \Delta\theta \cdot \eta_{\text{GB}}^{\text{Equilibrium}} = \sqrt{2\kappa_\eta^2 \cdot \Delta f\left(\eta_{\text{GB}}^{\text{Equilibrium}}, c_{\text{GB}}^{\text{Equilibrium}}\right)}. \tag{15}$$

The two boundary conditions for the function $c(\eta)$ are:

$$c(\eta=1) = c_\infty, \tag{16}$$

where $c_\infty$ is the composition inside the bulk (grain), and

$$\left.\frac{dc}{d\eta}\right|_{\eta_{\text{GB}}^{\text{Equilibrium}}} = 0. \tag{17}$$

Subsequently, they developed a 3D graphical construction method to solve $\eta_{\text{GB}}^{\text{Equilibrium}}$ and $c_{\text{GB}}^{\text{Equilibrium}}$ simultaneously based on equations (15-17) [79].

Here, we simplify that approach to allow 2D graphical construction (to help illustrate the key underlying physical picture intuitively) by first adopting the $c(\eta)$ profile and $c_{\text{GB}}(\eta_{\text{GB}})$ that minimize $\Delta F\left(\eta_{\text{GB}}\right)$ for a given $\eta_{\text{GB}}$ in equation (12). Then, we can solve the equilibrium level of the GB order (or GB disorder $\eta'_{\text{GB}} \equiv 1-\eta_{\text{GB}}$) by finding $\eta_{\text{GB}}^{\text{Equilibrium}}$ (only one variable) to minimize the excess grand potential (interfacial energy) in equation (11) via differentiation:

$$2s \cdot \Delta\theta \cdot \eta_{\text{GB}}^{\text{Equilibrium}} = -\left.\frac{d(\Delta F)}{d\eta_{\text{GB}}}\right|_{\eta_{\text{GB}}^{\text{Equilibrium}}}. \tag{18}$$

By comparing equations (15) and (18), we know:

$$-\left.\frac{d(\Delta F)}{d\eta_{\text{GB}}}\right|_{\eta_{\text{GB}}^{\text{Equilibrium}}} = 2\sqrt{2\kappa_\eta^2 \cdot \Delta f\left(\eta_{\text{GB}}^{\text{Equilibrium}}, c_{\text{GB}}^{\text{Equilibrium}}\right)}, \tag{19}$$

where $c_{\text{GB}}^{\text{Equilibrium}}$ is a function of $\eta_{\text{GB}}^{\text{Equilibrium}}$ (that we can solve numerically via the minimization in equation (12)). This equation is similar to Tang, Carter, and Cannon's prior analysis of the unary GB [80], except that a more complex form of $\Delta F\left(\eta_{\text{GB}}\right)$ for a binary GB without an analytical expression. But $\Delta F(\eta_{GB})$ is well defined by the minimization in equation (12). Noting that equations (12) and (18) are mathematically equivalent to the equations (15-17) from the Tang-Carter-Cannon model [79]. Here, we rearrange the order of energy minimization to allow us to draw 2D sketches for illustrating disorder-order transitions in a binary alloy without losing the rigorousness.

In the current case, we do not have all parameters to evaluate the $Bi_2O_3$-doped ZnO system quantitatively. Thus, we first schematically show that oxygen reduction can induce a GB disorder-to-order transition by decreasing the coupling coefficient $s$, (if it can be decreased) and subsequently justify by DFT calculations that the parameter $s$ can be decreased by ~2.4× in the reduced GB.

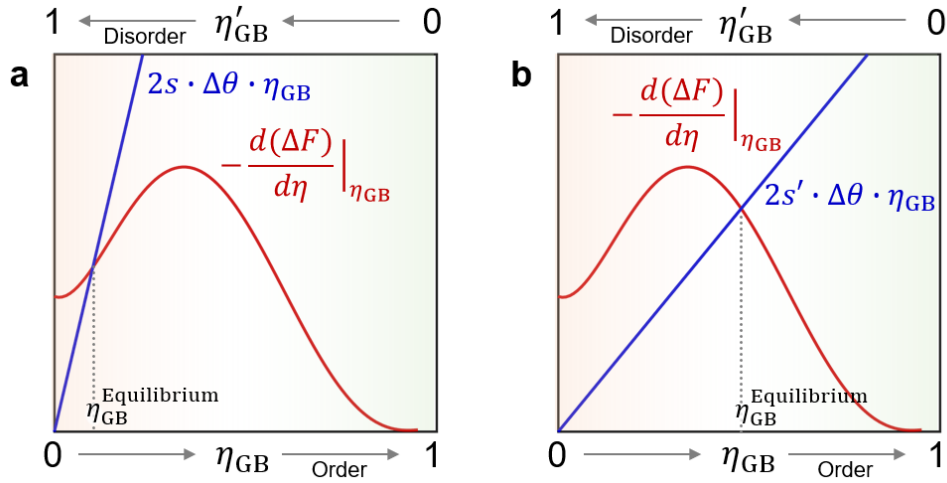

**Supplementary Fig. 17. Schematic illustration of the GB disorder-to-order transition using a generalizable thermodynamic model following Tang, Carter, and Cannon[79,81]. a**, Schematic illustration of a graphical construction method to solve the equation (2) in the main article or equation (18) here. The intersection of the red and blue lines indicates an equilibrium order parameter $\eta_{GB}^{Equilbrium}$. **b**, Decreasing the GB misorientation-disordering coupling parameter (*i.e.*, changing $s \to$ a smaller $s'$) can change the GB into an ordered GB state (with greater $\eta_{GB}^{Equilbrium}$). Since DFT calculations showed that the parameter $s$ for the reduced GB is substantially smaller than that for the stoichiometric GB (*i.e.*, decreasing $s$ by ~2.4× based on our DFT estimations), oxygen reduction can cause such a GB disorder-to-order transition. Thus, it explains the origin of the observed reduction-induced GB ordering in our experiments.

Equation (18) allows a 2D graphical construction to solve $\eta_{GB}^{Equilbrium}$ by plotting the both sides of the equation and finding the intersections. Equation (19) suggests that the right side of equation (18) (*i.e.*, the red curves in Supplementary Fig. 17) would likely to have two minima with a barrier in between because the function $\Delta f(c, \eta)$ has two minima for the liquid ($\eta = 0$) and solid ($\eta = 1$) phases. The Supplementary Fig. 17 schematically illustrates the GB disorder-order transition. The intersection of the red and blue lines in Supplementary Fig. 17a defines an equilibrium order parameter $\eta_{GB}^{Equilbrium}$ for a more disordered GB state. The Supplementary Fig. 17b shows that decreasing $s$ can change the GB into a more ordered state (with a greater $\eta_{GB}^{Equilbrium}$).

To support the above-proposed origin of reduction-induced GB order-to-disorder transition quantitatively, we used DFT calculations of the GB structures before and after relaxations (Supplementary Table 1) to estimate $s \cdot \Delta\theta$ for both the stoichiometric GB and the reduced GB. Before the DFT relaxation, we have:

$$\sigma^x(\eta_{GB}^{(0)}) = s \cdot \Delta\theta \cdot \left(\eta_{GB}^{(0)}\right)^2 + \Delta F\left(\eta_{GB}^{(0)}\right), \tag{20}$$

where $\eta_{GB}^{(0)}$ is the initial order parameter at center of GB ($x = 0$) before the DFT relaxation. After the relaxation, the equilibrium GB is obtained, for which we have:

$$\sigma^x(\eta_{GB}^{Equilibrium}) = s \cdot \Delta\theta \cdot \left(\eta_{GB}^{Equilibrium}\right)^2 + \Delta F\left(\eta_{GB}^{Equilibrium}\right), \tag{21}$$

We can use the DFT calculations to estimate $\Delta\sigma_{DFT}^x \equiv \sigma^x(\eta_{GB}^{(0)}) - \sigma^x(\eta_{GB}^{Equilibrium})$. Subsequently, combining equations (20) and (21) produces:

$$\Delta\sigma_{DFT}^x = s \cdot \Delta\theta \cdot \left[\left(\eta_{GB}^{(0)}\right)^2 - \left(\eta_{GB}^{Equilibrium}\right)^2\right] + \left[\Delta F\left(\eta_{GB}^{(0)}\right) - \Delta F\left(\eta_{GB}^{Equilibrium}\right)\right], \tag{22}$$

or:

$$s \cdot \Delta\theta \cdot \left[\left(\eta_{GB}^{(0)}\right)^2 - \left(\eta_{GB}^{Equilibrium}\right)^2\right] = \Delta\sigma_{DFT}^x + \left[\Delta F\left(\eta_{GB}^{Equilibrium}\right) - \Delta F\left(\eta_{GB}^{(0)}\right)\right]. \tag{23}$$

Noting that the first term in equation (22) dominates the relaxation, so we know:

$$s \cdot \Delta\theta \cdot \left[\left(\eta_{GB}^{(0)}\right)^2 - \left(\eta_{GB}^{Equilibrium}\right)^2\right] > \left[\Delta F\left(\eta_{GB}^{Equilibrium}\right) - \Delta F\left(\eta_{GB}^{(0)}\right)\right] > 0. \tag{24}$$

Thus, we can use a linear Taylor expansion to estimate the second term (that is smaller in its absolute value in comparison with the first term) in equation (22) to the first order of approximation, as:

$$\left[\Delta F\left(\eta_{GB}^{(0)}\right) - \Delta F\left(\eta_{GB}^{Equilibrium}\right)\right] \approx \left.\frac{d(\Delta F)}{d\eta_{GB}}\right|_{\eta_{GB}^{Equilibrium}} \cdot \left(\eta_{GB}^{(0)} - \eta_{GB}^{Equilibrium}\right). \tag{25}$$

Combining with equation (22), we can obtain:

$$\left[\Delta F\left(\eta_{GB}^{(0)}\right) - \Delta F\left(\eta_{GB}^{Equilibrium}\right)\right] \approx -2s \cdot \Delta\theta \cdot \eta_{GB}^{Equilibrium} \cdot \left(\eta_{GB}^{(0)} - \eta_{GB}^{Equilibrium}\right). \tag{26}$$

Plugging the above equation into equation (22), we have:

$$\Delta\sigma_{DFT}^x \approx s \cdot \Delta\theta \cdot \left[\left(\eta_{GB}^{(0)}\right)^2 - \left(\eta_{GB}^{Equilibrium}\right)^2\right] - 2s \cdot \Delta\theta \cdot \eta_{GB}^{Equilibrium} \cdot \left(\eta_{GB}^{(0)} - \eta_{GB}^{Equilibrium}\right). \tag{27}$$

Hence, we can estimate the $s \cdot \Delta\theta$ (where $\Delta\theta$ is dimensionless constant) via:

$$s \cdot \Delta\theta \approx \frac{\Delta\sigma_{\text{DFT}}^{x}}{\left(\eta_{\text{GB}}^{(0)} - \eta_{\text{GB}}^{\text{Equilibrium}}\right)^{2}} \tag{28}$$

Based on the DFT calculations of the stoichiometric and reduced GBs before and after the relaxation, we computed $s \cdot \Delta\theta$ for both stoichiometric and reduced GBs. The key results are shown in Supplementary Table 1. Since $\Delta\theta$ is a constant (the GB misorientation), this calculation suggested that the reduction can decrease the coupling parameter/coefficient *s* by about 2.4 times. Thus, it quantitively justifies that reduction can induce a GB disorder-to-order transition based on Supplementary Fig. 17.

This generalizable thermodynamic model and DFT-based approach can also be useful for predicting trends for other systems in future studies.

We shall make several additional notes for the sake of completeness and rigor (albeit that they are not the focus of the discussion here). First, we expect the red curves in Supplementary Fig. 17 have two minima and one maximum (based on equation (19)), but we do not have an analytical expression (that has to be obtained numerically based on equation (12)); yet the sketches in Supplementary Fig. 17 can clearly illustrate the physical pictures. Second, if the blue line intersects with the red curve twice, it defines two (stable and metastable) GB states and a possible first-order GB transition [79,81], which is a scientifically interesting interfacial phenomenon but not the focus of the discussion here. Third, the blue curve can be nonlinear beyond the adoption of the simple function of $g(\eta) = \eta^{2}$ (to represent the Ready-Shockley type behavior). Nonetheless, the current approach and sketches in Supplementary Fig. 17, along with the DFT calculations shown in Supplementary Table 1, can effectively illustrate the physical origin of the reduction-induced GB disorder-to-order transition via decreasing the parameter *s*.

**Supplementary Table 1. DFT calculations of the GB structures before and after the relaxations to evaluate the change of *s* to justify the disorder-order transition.** $E_{relax}$ and $E_{unrelax}$ are the energies of ZnO GBs after and before the relaxation. $\Gamma_{\text{Disorder}}$ is the GB excess of disorder (listed here for reference). $\eta_{\text{GB}}^{(0)}$ and $\eta_{\text{GB}}^{\text{Equlibrium}}$ are the order parameter at the center of unrelaxed and relaxed GB (at *x* = 0). The parameter *s* is decreased by ~2.4X in the reduced GB, which can justify the reduction-induced GB disorder-to-order transition.

|  | Stoichiometric GB | Reduced GB |
|---|---|---|
| $E_{relax}$ (eV/atom) | -4.480 | -4.385 |
| $E_{unrelax}$ (eV/atom) | -3.718 | -4.060 |
| $\Delta\sigma_{\text{DFT}}^{x}$ (eV/nm²) | 218 | 42 |
| $\Gamma_{\text{Disorder}}^{\text{Equilibrium}}$ (nm⁻²) | 25 | 14 |
| $\Gamma_{\text{Disorder}}^{(0)}$ (nm⁻²) | 14 | 11 |
| $\eta_{\text{GB}}^{\text{Equilibirum}}$ | 0.10 | 0.18 |
| $\eta_{\text{GB}}^{(0)}$ | 0.35 | 0.35 |
| $s \cdot \Delta\theta$ (eV/nm²) | ~3.5×10³ | ~1.4×10³ |

## Supplementary Note 9:

**Mechanisms of enhanced kinetics in the reduced GB: insights from the DFT calculated differential charge densities and Bader charges**

To further explain the underlying mechanisms of the enhanced GB kinetics (diffusivities and mobilities) of the reduced GBs, we calculated differential charge density and Bader charge transfers of the Bi atoms for the stoichiometric GB *vs.* the reduced GB.

First, the DFT-calculated isosurfaces, where yellow regions represent charge accumulation and cyan regions represent charge depletion, indicate stronger charge transfer in the stoichiometric GB than that in the reduced GB (Supplementary Fig. 18a *vs.* 18b). By comparing the 2D averaged differential charge density profile along *z* direction (Supplementary Fig. 18c *vs.* 18d), the large oscillation peaks for the stoichiometric GB verified the stronger charge transfer. Since atoms typically form strong chemical bonding with strong charge transfer, this stronger charge transfer can explain the low GB diffusivities in, and the mobility of, the stoichiometric GBs (because of the stronger bonding or "pinning" effects).

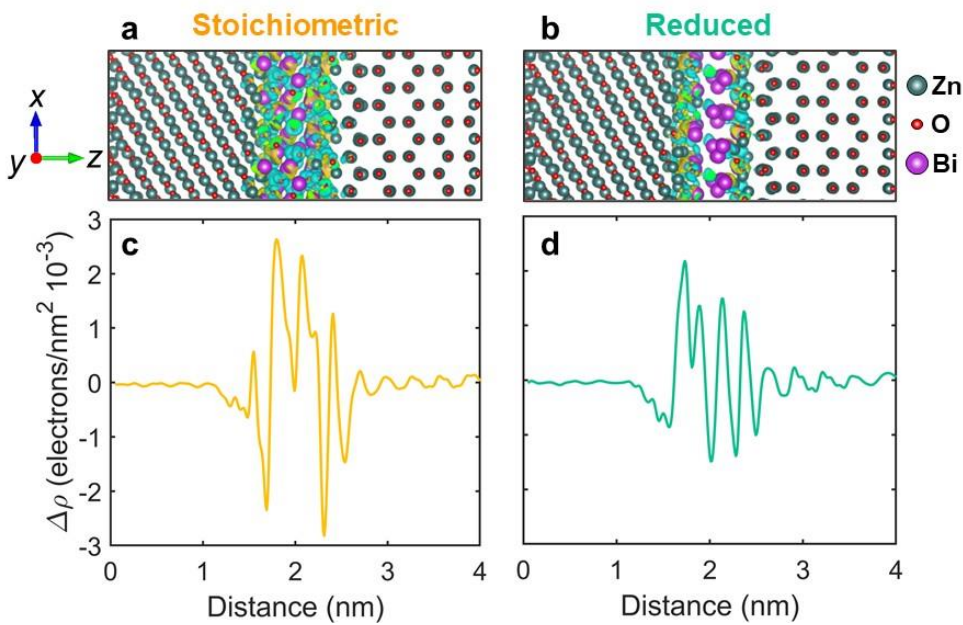

**Supplementary Fig. 18. Differential charge densities calculated from DFT to reveal the underlying mechanism of enhanced kinetics in the reduced GB.** Isosurfaces of the differential charge densities of (**a**) stoichiometric GB and (**b**) reduced GB. The isovalue is set to 0.01 e/Å$^3$ for the plots. The yellow and cyan isosurfaces, respectively, represent charge accumulation and depletion, respectively. The differential charge density profiles projected along the *z* direction of (**c**) stoichiometric GB and (**d**) reduced GB, suggesting weaker bonding in the reduced (ordered) GB

Second, the calculated average Bader charges showed that one Bi atom losses ~1.4 *e* in the stoichiometric GB, which is significantly greater than ~0.69 *e* in the reduced GB (Supplementary Table 2). Thus, the aliovalent Bi atoms serve as charged "hot spots" to provide more "pinning" effects in the stoichiometric GB, thereby decreasing the GB diffusivities and mobility. However, the oxygen reduction may significantly reduce the charged "hot spots" and associated strong bonding and "pinning" effects, thereby enhancing the GB diffusivities in, and the mobility of, the reduced GBs, which subsequently caused the enhanced and abnormal grain growth observed in experiments.

The above thermodynamic model and DFT results can be understood intuitively. On the one hand, the presence of aliovalent $Bi^{3+}$ adsorbates (substituting $Zn^{2+}$ cations with extra $O^{2-}$ for the charge compensation) in the stoichiometric GB will likely lead to interfacial disordering. On the other hand, the reduction will decrease the effective charge on Bi adsorbates (as shown by calculated Bader charges given in Supplementary Table 2) to make interfacial structure more ordered (with enhanced mobility).

**Supplementary Table 2. The average Bader charge transfer $\Delta q$ for the Bi atoms at the stoichiometric vs. reduced GB.** In comparison with the stoichiometric GB, the smaller average charge transfer for the Bi atoms at the reduced GB suggests weaker bonding, thereby explaining the increased kinetics of the reduced GB.

|  | Stoichiometric GB | Reduced GB |
| --- | --- | --- |
| $\Delta q_{Bi}$ (\|e\|) | 1.40 | 0.69 |

## Supplementary Note 10:

### Enhanced grain growth in reduced atmospheres supporting the proposed mechanism

To further support proposed mechanism that the reduction can enhance grain growth, we conducted a series of controlled grain growth experiments in air, Ar, and Ar + 5% $H_2$, respectively by isothermally annealing samples at 880 °C for 4 hours and quenching (Supplementary Fig. 19).

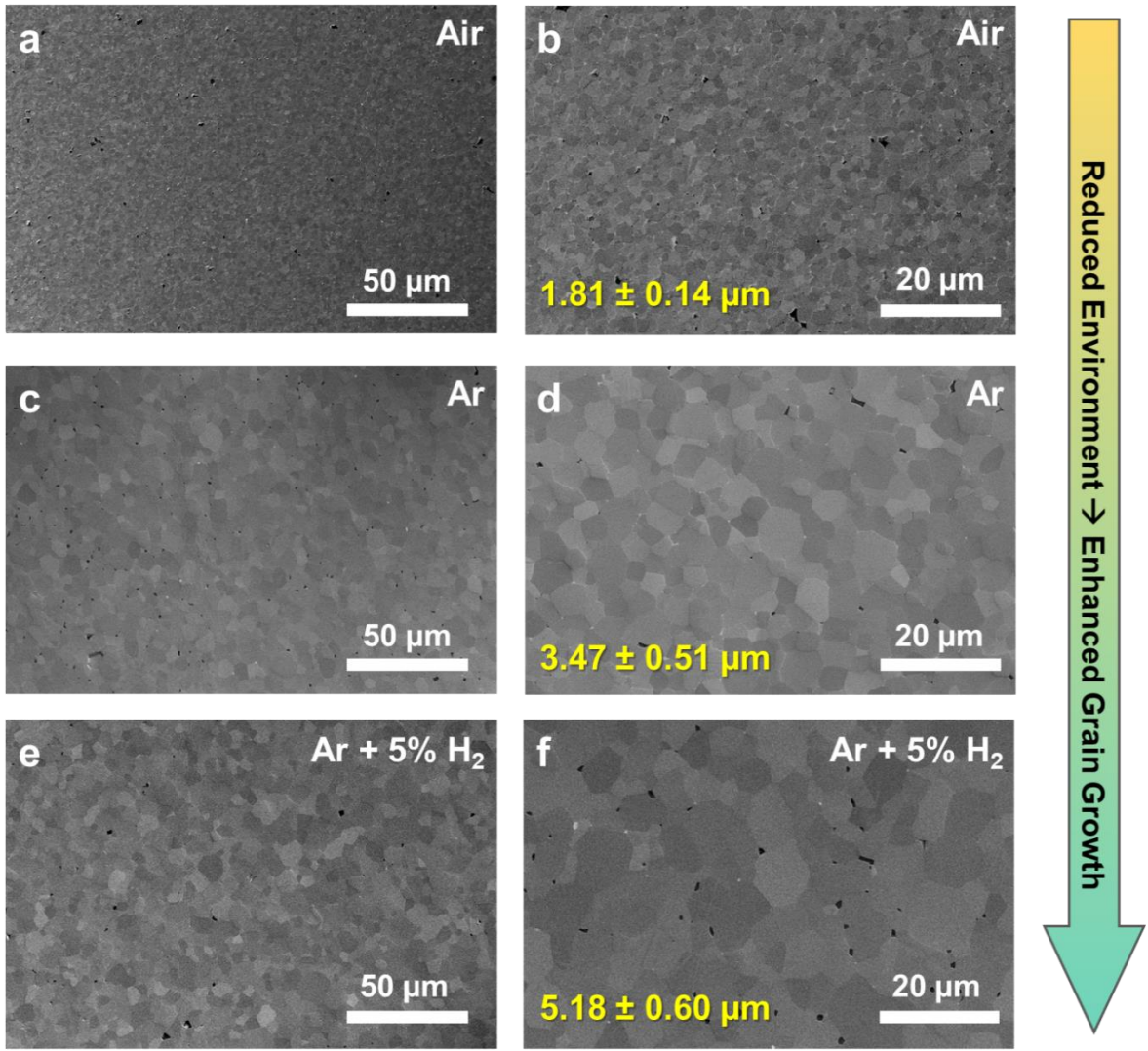

**Supplementary Fig. 19. Cross-sectional SEM micrographs of three ZnO + 0.5 mol% $Bi_2O_3$ specimens annealed in different atmospheres, showing that a reduced environment can promote grain growth (without an applied electric current).** Low- and high-magnification cross-sectional SEM micrographs of polycrystalline specimens quenched from 880 °C after annealing for 4 hours in (**a, b**) air, (**c, d**) Ar, and (**e, f**) in Ar + 5% $H_2$, respectively. The measured average grain sizes (± one standard deviations) are 1.81 ± 0.14 μm, 3.47 ± 0.51 μm, and 5.18 ± 0.60 μm, respectively.

The average grain size was measured to be 2.02 ± 0.17 µm for the specimen annealed in (the oxidized) air (Supplementary Fig. 19a). The measured grain size increased to 3.47 ± 0.51 µm for the specimen annealed in (the inert/reduced) Ar (Supplementary Fig. 19b), and it further increased to 5.18 ± 0.60 µm for the specimen annealed in the most reduced Ar + 5% $H_2$ atmosphere (Supplementary Fig. 19c).

Thus, this observation of enhanced grain growth in reduced atmospheres further supports our hypothesis that reduction can promote grain growth (even without an applied electric field/current), which are also supported by AIMD calculations of GB diffusivities (Fig. 6c).

## Supplementary Note 11:
### Electric characteristics of the specimens

During our isothermal annealing experiments, electric potentials and currents were recorded using a high-precision digital multimeter (Tektronix DMM 4050, Beaverton, Oregon, USA) during the experiments. The measured resistance *vs.* time curve for the PC/SC/PC sandwich specimen is shown in Supplementary Fig. 20b. For comparison, we also measured the resistivity *vs.* time curves for both a polycrystal and a single crystal (Supplementary Fig. 20a). While the resistances and resistivities changed with time initially due to evolution of defects and microstructures as well as polarization, they reached steady states after ~50 mins. At the steady state, the total resistance of the PC/SC/PC sandwich specimen was identical to that calculated from the measured resistivities of the polycrystal and the single crystal within the measurement errors.

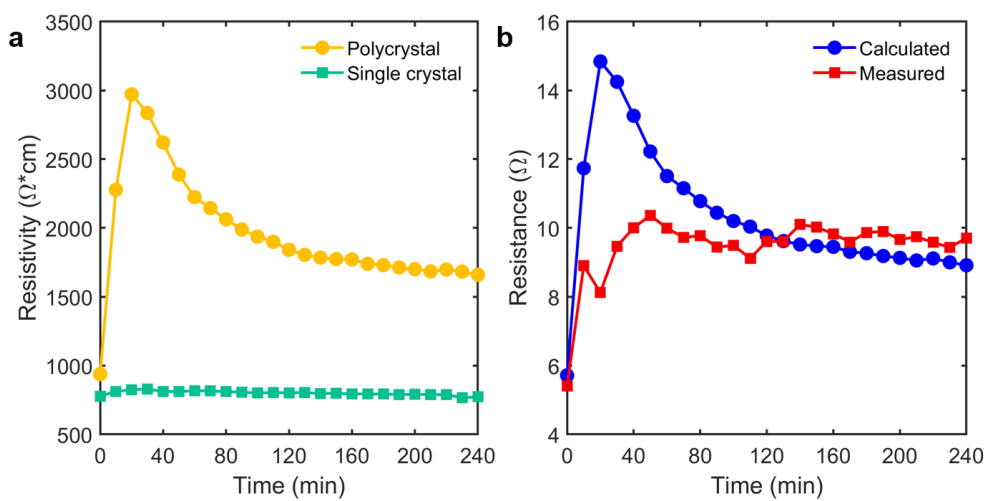

**Supplementary Fig. 20. The resistance *vs*. annealing time of a sandwich specimen, a single crystal, and a polycrystal specimen isothermally annealed at a furnace temperature of 840 °C with the identical constant current density $J$ = 6.4 mA/mm$^2$. a**, The measured resistivity of the Bi$_2$O$_3$-doped ZnO polycrystal and single crystal *vs*. annealing time. **b**, The measured resistance of a PC1/SC/PC2 sandwich specimen (characterized in this study) and the calculated resistance (from the measured resistivities of the single crystal and polycrystal shown in Panel (a) and the geometry of the sandwich specimen) *vs*. annealing time. The thicknesses of PC1, SC, and PC2 of this specimen are 0.565, 0.5, and 0.565 mm, respectively. After reaching a steady state at ~50 minutes, the measured and calculated resistances of the sandwich specimens largely agree with each other. The estimated steady-state electric fields are 9.2 V/cm in the polycrystal sections (PC1 and PC2) and 9.9 V/cm for single crystal (SC) section, respectively, of the sandwich specimen.

The steady-state electric fields were calculated to be ~9.2 V/cm in the polycrystal sections (PC1 and PC2) and ~9.9 V/cm for single crystal (SC) section, respectively, of the sandwich specimen.

# Supplementary References: